\documentclass[12pt]{article}

\usepackage{amsmath}
\usepackage{amsfonts}
\usepackage{mathrsfs}
\usepackage{natbib}
\usepackage{comment}
\usepackage{bm}
\usepackage{changes}
\usepackage{booktabs}
\usepackage{courier}
\usepackage{graphicx}
\usepackage{subcaption}

\newcommand{\E}{{\mbox{E}}}
\newcommand{\VAR}{{\mbox{VAR}}}
\newcommand{\COV}{{\mbox{COV}}}

\newcommand{\x}{{\bm{\mathrm{x}}}}
\newcommand{\fx}{{\bm{\mathrm{f}}_X}}
\newcommand{\fy}{{\bm{\mathrm{f}}_Y}}
\newcommand{\hfx}{{\widehat{\bm{\mathrm{f}}}_X}}
\newcommand{\hfy}{{\widehat{\bm{\mathrm{f}}}_Y}}
\newcommand{\freg}{{\widehat{\bm{\mathrm{f}}}_\mathrm{reg}}}

\newcommand{\C}{{\bm{\mathrm{C}}}}
\newcommand{\D}{{\bm{\mathrm{D}}}}
\newcommand{\1}{{\bm{\mathrm{1}}}}
\newcommand{\0}{{\bm{\mathrm{0}}}}
\newcommand{\A}{{\bm{\mathrm{A}}}}
\newcommand{\B}{{\bm{\mathrm{B}}}}
\newcommand{\bb}{{\bm{\mathrm{b}}}}
\newcommand{\bS}{{\bm{\mathrm{\Sigma}}}}

\newcommand{\hfxl}{{\widehat{\bm{\mathrm{f}}}_{X,\lambda}}}
\newcommand{\hfyi}{{\widehat{f}_{Y,i}}}
\newcommand{\hfyj}{{\widehat{f}_{Y,j}}}

\newcommand{\lmds}{{\lambda_{\mathrm{SURE}}}}
\newcommand{\lmdsr}{{\lambda_{\mathrm{Scree}}}}

\newcommand{\argmin}{{\rm{argmin}}}

\newcommand{\blind}{0}

\usepackage{times}

\begin{document}

\def\spacingset#1{\renewcommand{\baselinestretch}%
{#1}\small\normalsize} \spacingset{1}


\if0\blind
{
  \title{\bf Density Deconvolution with Additive Measurement Errors using Quadratic Programming}

      \author{Ran Yang, Daniel Apley\thanks{Corresponding author. Department of Industrial Engineering \& Management Sciences, Northwestern University, Evanston, Illinois,60208-3119}, Jeremy Staum, 
      David Ruppert\thanks{Department of Statistical Science and School of Operations Research and Information Engineering, Cornell University, Ithaca, NY 14853-3801}
   }
  \maketitle
} \fi

\if1\blind
{
  \bigskip
  \bigskip
  \bigskip
  \begin{center}
    {\LARGE\bf Density Deconvolution using Quadratic Programming}
\end{center}
  \medskip
} \fi

\bigskip
	
\begin{abstract}
Distribution estimation for noisy data via density deconvolution is a notoriously difficult problem for typical noise distributions like Gaussian. We develop a density deconvolution estimator based on quadratic programming (QP) that can achieve better estimation than kernel density deconvolution methods. The QP approach appears to have a more favorable regularization tradeoff between oversmoothing vs. oscillation, especially at the tails of the distribution. An additional advantage is that it is straightforward to incorporate a number of common density constraints such as nonnegativity, integration-to-one, unimodality, tail convexity, tail monotonicity, and support constraints. We demonstrate that the QP approach has outstanding estimation performance relative to existing methods. Its performance is superior when only the universally applicable nonnegativity and integration-to-one constraints are incorporated, and incorporating additional common constraints when applicable (e.g., nonnegative support, unimodality, tail monotonicity or convexity, etc.) can further substantially improve the estimation.   
\end{abstract}
		
\vfill

\newpage
\spacingset{1.45} 

\section{Introduction}\label{sec:introduction}
We consider the following statistical problem. Suppose a random variable (r.v.) $X$ and its probability density function (pdf) $f_X(\cdot)$, cumulative distribution function (cdf) $F_X(\cdot)$, and various quantiles are of interest, but only a random sample of noisy observations $\{Y_1,\cdots,Y_n\}$ are available with which to estimate the pdf, cdf, and quantiles. The underlying model is $Y_i=X_i+Z_i  ,i\in\{1,2,\cdots,n\}$, where the $Z_i$'s represent observation errors and are independent of the $X_i$'s. As is typical in the extensive literature on density estimation with noisy observations, (e.g., \cite{carroll1988optimal,stefanski1990rates, fan1991optimal,diggle1993fourier,delaigle2004practical,hall2007ridge,meister2009deconvolution}), the pdf $f_Z$ of $Z$ is assumed to be known. Existing estimators for this problem have slow convergence rates and poor finite-sample accuracy. Although their asymptotic convergence rates are optimal and thus cannot be improved, in this paper we propose new estimates based on quadratic programming whose finite-sample performance improves over existing estimators substantially. We note that most of the prior work on this topic casts the problem directly in terms of pdf estimation and refers to it as density deconvolution, recognizing that estimates of the cdf and the quantiles can be obtained in the obvious manner from an estimate of the pdf. We adopt the same convention in this paper, although we are interested in cdf and quantile estimation, in addition to pdf estimation. 

For the additive measurement error model, the pdf $f_Y$ of $Y$ is the convolution
\begin{equation}\label{conveq}
f_Y(y)=\left(f_X*f_Z\right)(y)=\int_{-\infty}^\infty f_Z(y-x)f_X(x)\mathrm{d}x                        
\end{equation}
This convolution in the spatial domain corresponds to multiplication $\phi_Y(\omega)=\phi_X(\omega)\cdot\phi_Z(\omega)$ in the Fourier domain, where $\phi_Y$ denotes the Fourier transform of $f_Y$ (likewise for $\phi_Z$ and $\phi_X$), and $\omega$ denotes frequency. In light of this, one classic and popular method is the Fourier-based kernel deconvolution (KD) (e.g., \cite{carroll1988optimal}, \cite{stefanski1990deconvolving}, \cite{diggle1993fourier}). One estimates $f_X(\cdot)$ as the inverse Fourier transform of $\phi_K (h\omega) \widehat{\phi}_Y(\omega)/\phi_Z(\omega)$ (the overscore symbol $\hat{\cdot}$ denotes an estimate). The additional term $\phi_K (h\omega)$ is a frequency-domain kernel weighting function that gives less weight to higher frequency values in the Fourier inversion integral to avoid numerical conditioning problems, and $h$ here is the bandwidth parameter for kernel smoothing. This approach is referred to as KD, because it is equivalent to kernel density estimation in the spatial domain, where the spatial domain kernel is the inverse Fourier transform of $\phi_K (h\omega) /\phi_Z(\omega)$, instead of some standard (e.g., Gaussian) kernel. Thus, KD is related to kernel density estimation for data observed without error \citep{rosenblatt1956remarks,parzen1962estimation,silverman1986density}.

Although KD methods have a sound theoretical foundation with well-understood asymptotic properties, their performance is sensitive to choice of $\phi_K(\cdot)$ and its bandwidth parameter that dictates the amount of smoothing (\cite{fan1991optimal}; \cite{barry1995choosing}; \cite{delaigle2004practical}), and it may be difficult to achieve a desirable balance between over- and under-smoothing, as illustrated in the example below. Moreover, methods having desirable asymptotic results do not necessarily perform well in typical finite sample situations. Other existing methodologies for density estimation include the spline-based smoothing method by \cite{silverman1984spline}, the wavelet-based method by \cite{pensky1999estimation}, and the wavelet-like method by \cite{comte2006penalized}. The smoothing splines methods are similar to KD in that they correspond approximately to smoothing by a kernel method with bandwidth depending locally rather than globally on the design points. Hence, such spline-based methods can suffer from similar issues with KD methods. Wavelet-based density deconvolution methods often have advantages over traditional KD methods for pdfs that have discontinuities and sharp peaks, but they can sometimes perform poorly for smooth functions. 

\begin{figure}[!tbp]
	\centering
	\subcaptionbox{\label{fig:intro1}}{\includegraphics[width=0.65\textwidth]{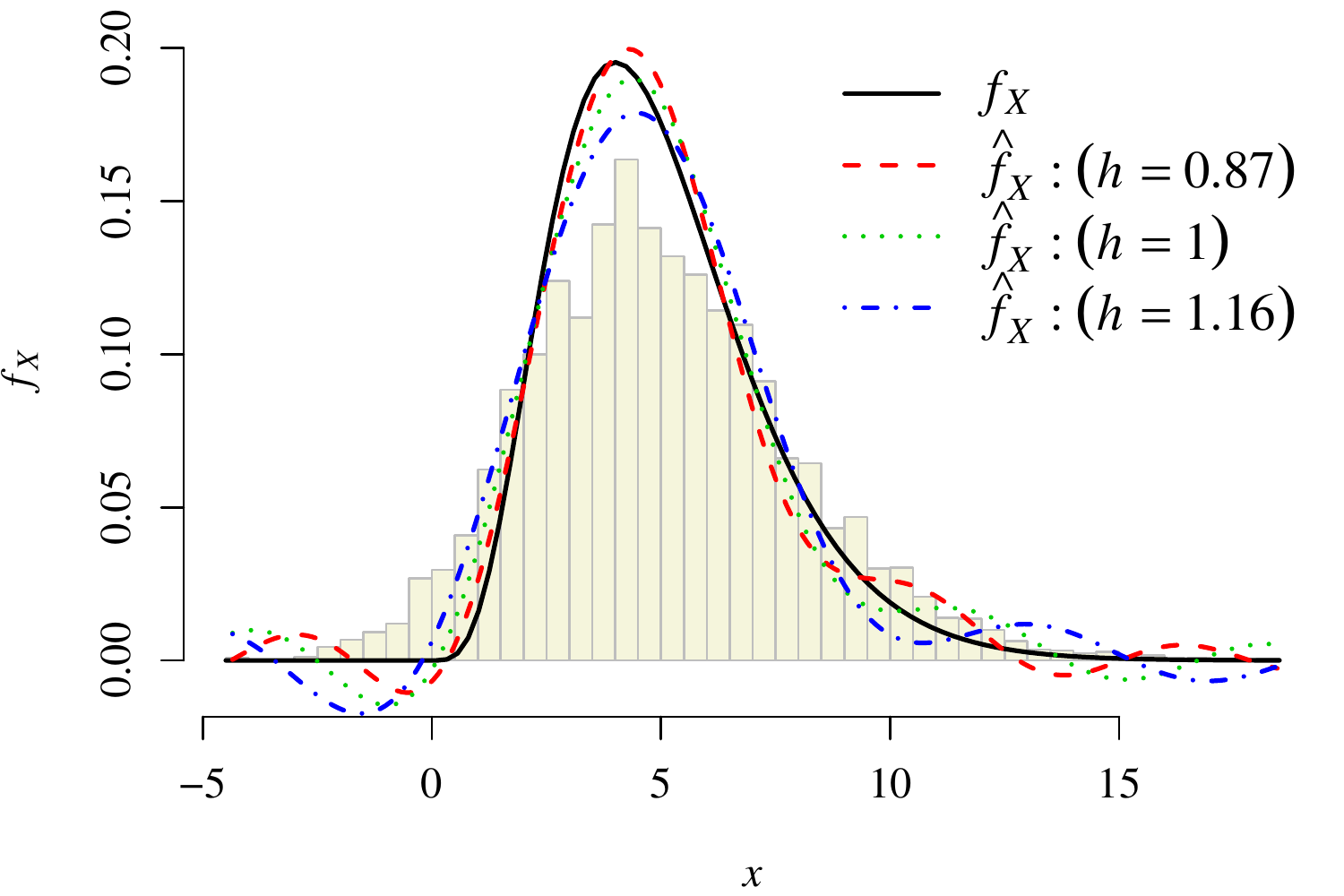}}
	\hfill
	\subcaptionbox{\label{fig:intro2}}{\includegraphics[width=0.65\textwidth]{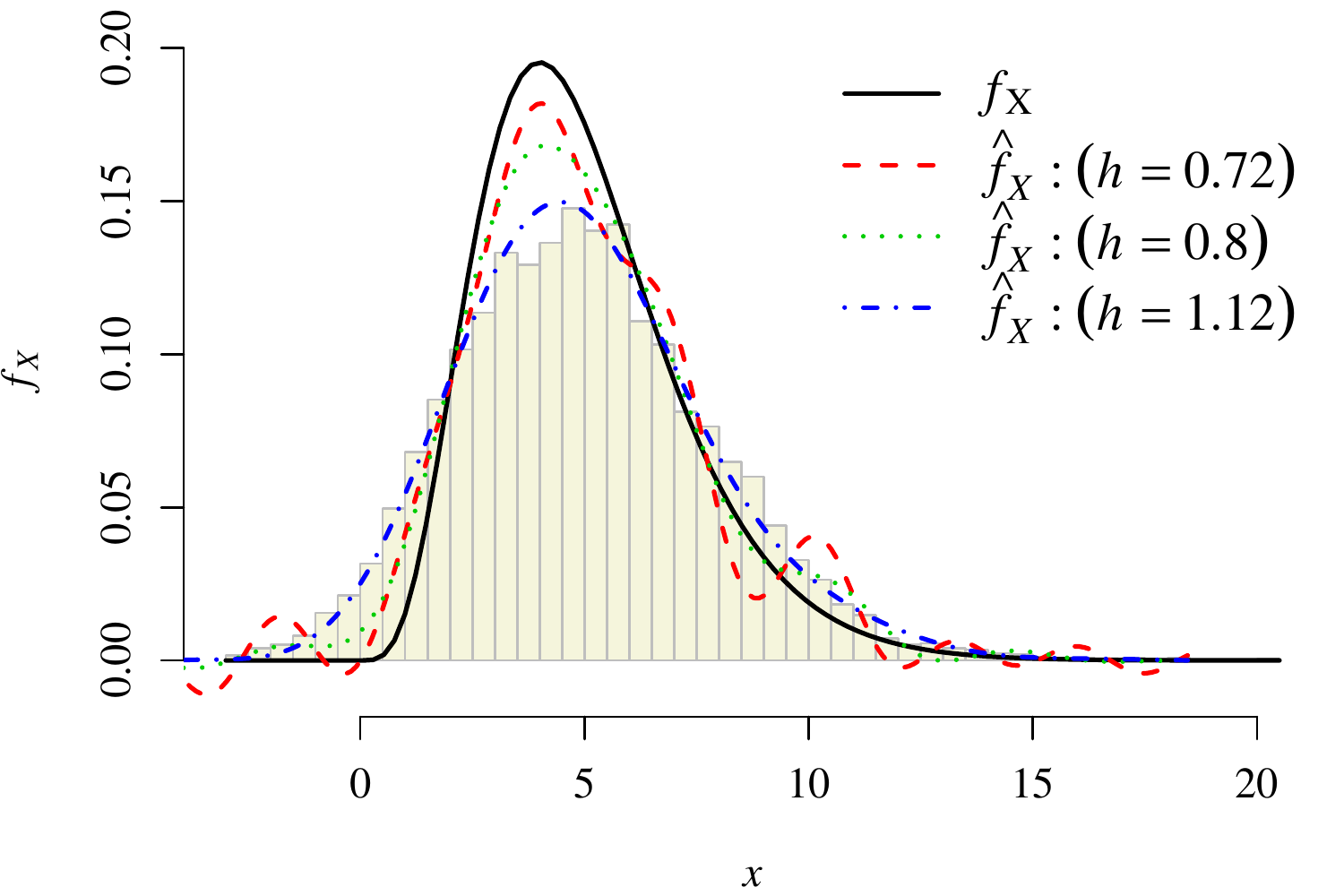}}
	\caption{The histogram is of $\{Y_1,\cdots,Y_n\}$ along with KD results for the $Gamma(5,1)$ example for various levels of smoothing bandwidth $h$ using (a) a rectangular kernel $\phi_K(\omega)=I_{[-1,1]}(\omega)$ and (b) a triweight kernel $\phi_K(\omega)=(1-\omega^2 )^3I_{[-1,1]}(\omega)$. Small $h$ corresponds to undersmoothing, and large $h$ corresponds to oversmoothing.}\label{fig:introplot}
\end{figure}

Fig.~\ref{fig:introplot} illustrates the performance of KD methods with two types of kernels for a gamma example in which $X \sim Gamma(5,1)$ (5 is the shape parameter and 1 is the rate parameter), $Z\sim N(0,\sigma_Z^2=3.2)$, and $n=5000$. A histogram of the observed data $\{Y_1,\cdots,Y_n\}$, along with the true density $f_X(\cdot)$, are shown in each panel. Panel (a) also shows the KD estimate $\widehat{f}_X$ with rectangular frequency domain kernel $\phi_K (\omega)=I_{[-1,1]}(\omega)$ for bandwidth parameter $h\in\{0.87,1.0,1.16\}$. The salient characteristic here is the pronounced oscillation on the tails of $\widehat{f}_X$. This oscillation can be reduced by increasing $h$, but the downside of this is oversmoothing of $\widehat{f}_X$. Even the largest $h=1.16$ has not eliminated the tail oscillation, and yet the peak of $f_X(\cdot)$ is already being oversmoothed. Panel (b) shows similar results, but for triweight kernel $\phi_K(\omega)=(1-\omega^2 )^3 I_{[-1,1]}(\omega)$. The same problematic tradeoff regarding the choice of bandwidth parameter is evident: If we choose a large enough bandwidth to avoid tail oscillation, this causes oversmoothing; and if we choose a small enough bandwidth to avoid oversmoothing, this causes tail oscillation. There may exist no value of bandwidth parameter that mitigates the tail oscillation without oversmoothing peaks.

Another undesirable characteristic of the KD method is that $\widehat{f}_X$ may be negative, as can be seen in Fig.~\ref{fig:introplot}. One can easily add a postprocessing adjustment of $\widehat{f}_X$ so that it is nonnegative and integrates-to-one, but this generally does not improve overall measures of quality of the estimator. As we demonstrate later, it is much more effective to incorporate these constraints more directly into the estimation process, as we do in our proposed estimator. Moreover, it is even more difficult to incorporate more complex shape constraints (e.g., tail monotonicity or convexity, unimodality, etc.) into the KD method. In contrast, it is straightforward to incorporate such shape constraints into our approach, when such knowledge is available, which we also demonstrate improves performance.

Motivated by the preceding, we develop a quadratic programming (QP) optimization approach for density deconvolution, together with an accompanying \textbf{R} package \textbf{QPdecon}. Specifically, our QP estimator $\widehat{f}_X$ is chosen to minimize a quadratic objective function that measures the difference between the convolution $\widehat{f}_X*f_Z$ and an empirical density estimator $\widehat{f}_Y$. A variety of shape constraints are translated into linear and convex constraints and can be easily incorporated into our QP formulation. In our objective function, we also include a quadratic regularization penalty for the purpose of ensuring the most appropriate level of smoothing. In order to select the regularization parameter (analogous to the bandwidth parameter in the KD method) we develop a simple and computationally efficient method based on a concept similar to Stein's unbiased risk estimator (SURE), which originates from \cite{mallows1973some}, \cite{stein1981estimation} and \cite{efron1986biased}.

Our examples indicate that, even without shape constraints, our QP estimator performs substantially better than both the classic KD method implemented with our own codes and the newer wavelet-like penalized contrast (PC) method (\cite{comte2006penalized}) implemented by the \textbf{R} package \textbf{deamer} (\cite{deamer}), which is the best performing existing package we have found so far. With shape constraints (when applicable), the performance improvement is even larger. Even when the error density $f_Z$ is Gaussian, which is notoriously difficult to deconvolve because of its smoothness (\cite{carroll1988optimal}; \cite{stefanski1990rates}; \cite{stefanski1990deconvolving}; \cite{fan1992deconvolution}; \cite{wang2011deconvolution}), our QP estimator can achieve reasonable performance. The conclusion that performance can be improved when appropriate shape constraints are incorporated is consistent with findings in the large body of prior work that has incorporated shape constraints in density estimation with error-free data, e.g., \cite{turnbull2014unimodal}, \cite{zhang1990fourier}, \cite{dupavcova1992epi}, \cite{papp2014shape}, \cite{royset2013nonparametric}, and in the limited prior work that has incorporated shape constraints in KD (\cite{carroll2011testing}; \cite{birke2009shape}). Although there are no proofs of asymptotic performance for the QP method, our focus is on density deconvolution in finite-sample situations, and we demonstrate that our proposed method works well via numerical studies on a variety of examples.

Optimization criteria like the quadratic objective function that we use in our QP estimator are much more amenable to incorporating shape constraints than other density deconvolution approaches. Optimization-based estimators using a regularized version of likelihood (\cite{staudenmayer2008density}; \cite{lee2013deconvolution}) or least squares (\cite{lee2015least}) as the objective function were recently considered for density deconvolution, although these works did not investigate the effects of incorporating shape constraints, as we do in this work. Another difference between our work and \cite{lee2015least} is that we derive a computationally efficient SURE-like approach for selecting the most appropriate value for the regularization parameter, whereas \cite{lee2015least} used the simulation-based approach of \cite{lee2013deconvolution}. We also introduce a simple graphical method that serves as a check on the selected regularization parameter, and we demonstrate that it is effective at preventing poor estimation results in the small proportion of cases where the SURE-like method selects the regularization parameter that results in too little regularization.

The remainder of the article is organized as follows. Section \ref{sec:QPapproach} describes our quadratic programming (QP) objective function for the density deconvolution problem (Section \ref{sec:basicQP}) and how to represent various shape constraints as linear constraints in the QP optimization (Section \ref{sec:shapeQP}). Section \ref{sec:selection} first derives the SURE-like method for selecting the regularization parameter and method of regularization (Section \ref{sec:SURE}) and then develops the simple, yet effective graphical check on the selected value (Section \ref{sec:screeplot}). Section \ref{sec:discussion} uses simulation examples to demonstrate the superior estimation performance of the QP approach, relative to the KD and PC approaches. We also discuss the effects of incorporating shape constraints on QP estimator performance and the performance of the SURE-like approach and the graphical check for selecting the regularization parameter. Section \ref{sec:conclusion} concludes the paper.

\section{QP approach for density deconvolution}\label{sec:QPapproach}
\subsection{Basic QP Problem Formulation}\label{sec:basicQP}
\begin{figure}[!tbp]
	\centering
	\includegraphics[width=0.6\textwidth]{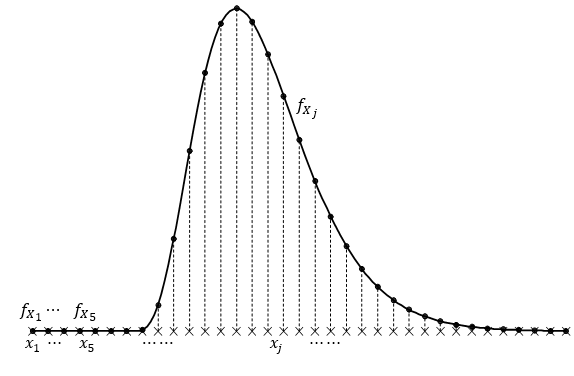}
	\caption{Illustration of the discrete approximation of $f_X$ and the notation. The black solid curve is the density of $f_X$; the black dots are the discretized approximation of $f_X$; and the black crosses are the $x$-locations for the discretization.\label{fig:illustration}}
\end{figure}

In the QP approach, we work with a discretized version of the continuous convolution in Eq.~(\ref{conveq}) over a grid of equally spaced points $\x=\{x_j:1\leq j\leq K\}$ for $f_X(\cdot)$ and $f_Y(\cdot)$, where $x_1=\min\{Y_i: 1\leq i \leq n\}$, and $x_K=\max\{Y_i: 1\leq i\leq n\}$. More specifically, defining $\delta=(x_K-x_1)/(K-1)$, we use the discrete approximation

\begin{equation*}
f_X(x)\cong f_X(x_j )\equiv f_{X,j},\text{if } x\in[x_j-\delta/2,x_j+\delta/2), \text{for } 1\leq j \leq K,
\end{equation*}
and similarly for $f_Y(\cdot)$, as illustrated in Fig.~\ref{fig:illustration}. Let the vectors $\fx=[f_{X,1},f_{X,2},\cdots,f_{X,K}]^T$ and $\fy=[f_{Y,1},f_{Y,2},\cdots,f_{Y,K}]^T$ represent the pdfs $f_X(\cdot)$ and $f_Y(\cdot)$, respectively. As an estimate of $\fy$, we will use the histogram of $\{Y_1,\cdots,Y_n\}$ with bins centered at the same set of support points $\x$. That is, the estimate $\widehat{f}_{Y,j}$ of $f_{Y,j}$ is the histogram bin height at $x_j$. Our discretized estimator $\hfx$ of the pdf $f_X(\cdot)$ will also be represented as a $K$-length vector. It should be noted that the QP approach inherently produces a smoothed estimate $\hfx$, so that further smoothing is unnecessary. Guidelines for selecting $K$ are discussed in Section \ref{sec:SURE}.

The discretized version of Eq.~(\ref{conveq}) can be written as
\begin{eqnarray}\label{dconveq}
\fy\cong\C\fx\Longleftrightarrow
\begin{bmatrix}
f_{Y,1}\\
\vdots\\
f_{Y,K}
\end{bmatrix}\cong\delta
\begin{bmatrix}
f_Z(x_1-x_1)&\cdots&f_Z(x_1-x_K)\\
\vdots&\ddots&\vdots\\
f_Z(x_K-x_1)&\cdots&f_Z(x_K-x_K)
\end{bmatrix}
\begin{bmatrix}
f_{X,1}\\
\vdots\\
f_{X,K}
\end{bmatrix},
\end{eqnarray}				
where the elements of the convolution matrix $\C$ are determined from the noise distribution, which is assumed known. At first glance, one may be tempted to use the estimate $\hfx=\C^{-1}\hfy$, which is an exact solution to Eq.~(\ref{dconveq}) with $\fx$ and $\fy$ replaced by their estimates. However, as is well known in the deconvolution literature, $\C$ is typically (for typical noise distributions) so poorly conditioned that $\C^{-1}\hfy$ is an unusable estimator subject to wild high-frequency oscillations.

Noting that $\C^{-1}\hfy$ is the solution to $\hfx=\argmin_{\fx} \lVert \hfy-\C\fx\rVert^2$, this suggests using the estimator
\begin{equation}\label{QP}
\hfx=\argmin_{\fx}\lVert \hfy-\C\fx\rVert^2+\lambda Q(\fx),						
\end{equation}
where $Q(\fx)$ is a regularization term that penalizes an $\fx$ that is poorly behaved in some respect, and $\lambda$ is a regularization parameter to be selected based on the data. For example, penalizing a large second derivative of $f_X(\cdot)$ can be achieved by using $Q(\fx)=\lVert\D_2\fx\rVert^2$, where $\D_2$ is an appropriately defined second-order difference matrix operator. We refer to this as second derivative regularization. Another option is to use $Q(\fx)=\lVert\fx-\freg\rVert^2$ where $\freg$ is some easily-determined and well-behaved approximation to $\fx$. In our examples, we take $\freg$ to be a Gaussian distribution with mean $\widehat{\mu}_Y$ and variance $\widehat{\sigma}_Y^2-\sigma_Z^2$, where $\widehat{\mu}_Y$ and $\widehat{\sigma}_Y^2$ are the sample mean and variance of $\{Y_1,\cdots,Y_n\}$. We refer to this as Gaussian regularization. Based on our simulation studies, the two regularization approaches performed comparably overall, with one method working better for some examples, and vice-versa for other examples. In Section \ref{sec:selection}, we present approaches for choosing the regularization method $Q(\fx)$, as well as $\lambda$.

Because all pdfs integrate-to-one and are nonnegative, it makes sense to incorporate this knowledge into the estimation of $\fx$ by including constraints in the QP formulation:
\begin{eqnarray}\label{QP0}
\hfx&=&\argmin_{\fx}\lVert \hfy-\C\fx\rVert^2+\lambda Q(\fx)\nonumber\\
s.t.& &\delta\1^T \fx=1\nonumber\\
& & \fx \geq \0 ,  								
\end{eqnarray}
where $\1$ is a column vector of ones, and $\fx\geq\0$ means that all elements of $\fx$ are nonnegative. 
Regarding quantifying the sampling variability in the QP deconvolution estimator, if desired, bootstrapping methods could be used

\subsection{Additional Shape Constraints}\label{sec:shapeQP}
One might have prior knowledge of various constraints on the shape of $f_X(x)$, e.g., that it is unimodal or that it has only nonnegative support. In this section, we discuss a number of such constraints that are common and that can be conveniently represented via linear constraints in the QP formulation (\ref{QP0}). Here, ``linear constraint" means that a linear transformation of the vector $\fx$ satisfies some specified equality or inequality constraint and not that $f_X(x)$ is constrained to be a linear function of $x$. It is intuitively reasonable to suppose that including any such prior knowledge of shape constraints will improve the estimation, and in Section \ref{sec:discussion} we demonstrate that this is indeed the case.

\textbf{\textsl{Tail monotonicity}}. Many pdfs have nonincreasing right tails and/or nondecreasing left tails. Suppose we know that $f_X(x)$ is nonincreasing for $x\geq x^m$ for some specified $x^m\in\x$. This can be handled by incorporating additional inequality constraints into the QP formulation (\ref{QP0}), as follows.
\begin{eqnarray*}
	\hfx&=&\argmin_{\fx}\lVert \hfy-\C\fx\rVert^2+\lambda Q(\fx)\\
	s.t.& &\delta\1^T \fx=1\\
	& & \fx \geq \0\\
	& & \A_m \fx\geq \0 ,   								
\end{eqnarray*}		
where
\[
\A_m=
\begin{bmatrix}
0&\cdots&0&1&-1&0&0&\cdots&0\\
0&\cdots&0&0&1&-1&0&\cdots&0\\
\vdots& &\vdots&\ddots& &\ddots& &\ddots&\vdots\\
0&\cdots&0&&\cdots&0&1&-1&0\\
0&\cdots&0&0&\cdots&0&0&1&-1
\end{bmatrix},
\]
and the first non-zero column of $\A_m$ corresponds to $x^m$. A nondecreasing left tail can be handled in a similar manner, by augmenting $\A_m$ with additional rows.

\textbf{\textsl{Tail convexity}}. Many pdfs also have one or both tails that are convex. Suppose we know that $f_X(x)$ is convex for $x\geq x^c$ for some specified $x^c\in\x$. This can be handled by adding the inequality constraints $\A_c\fx\geq\0$, where
\[
\A_c=\begin{bmatrix}
0&\cdots&0&1&-2&1&0&0&\cdots&0\\
0&\cdots&0&0&1&-2&1&0&\cdots&0\\
\vdots& &\vdots&  &\ddots& &\ddots& &\ddots&\vdots\\
0&\cdots&0&\cdots&&0&1&-2&1&0\\
0&\cdots&0&0&\cdots&0&0&1&-2&1
\end{bmatrix},
\]
and the first non-zero column of $\A_c$ corresponds to the location of $x^c\in\x$. A convex left tail can be handled similarly. We impose the convexity constraints as linear inequality constraints in the QP formulation and only for the tails of the pdf, not more general convex constraints to the entire pdf. 

\textbf{\textsl{Unimodality}}. If we know the pdf is unimodal with mode at known location $x^u\in\x$, this is equivalent to a nonincreasing monotonicity constraint for $x\geq x^u$ and a nondecreasing monotonicity constraint for $x\leq x^u$. In analogy with the form of the monotonicity constraint given earlier, this can be handled by adding the inequality constraints $\A_u\fx\geq \0$, where
\[
\A_u=\begin{bmatrix}
-1&1&0&\cdots& &\cdots&0\\
0&\ddots&\ddots&\ddots&&\ddots &\vdots\\
\vdots&\ddots&-1&1 &0&\cdots &0\\
0&\cdots&0&1 &-1&\ddots &0\\
\vdots&\ddots&&\ddots& &\ddots&0\\
0& &\cdots& &0&1 &-1
\end{bmatrix},
\]
and the row of $\A_u$ in which the order of the elements transitions from $\{-1,1\}$ to $\{1,-1\}$ corresponds to the mode location $x^u$. The preceding is relevant when the mode location $x^u$ is known in advance, which generally will not be the case. For unknown mode locations, one can add $x^u$ as an additional decision variable and solve $K$ separate QPs, each with a different unimodality constraint corresponding to each candidate $x^u\in\x$. The value of $x^u$ resulting in the smallest QP objective function value would be concluded the mode location.

\textbf{\textsl{Support constraints}}. If there is information on the support of $f_X(x)$, e.g., that $f_X(x)=0$ for $x<0$, this can be easily taken into account. As an example, suppose that $X\geq 0$ is the concentration of a trace impurity in a chemical production process, and $Y$ is a noisy measurement of $X$ that can assume negative values, even though $X$ is nonnegative. In situations like this, it is reasonable to suppose that we can improve our estimate $\hfx$ by taking into account the information that $f_X(x)=0$ over certain regions, even though $f_Y(x)>0$ over these regions. To handle this, supposing we know that the support of $f_X(x)$ lies within the interval $[x_a,x_b]$ for some specified $x_1\leq x_a<x_b\leq x_K$, one could solve (\ref{QP0}) with the additional constraints that $f_{X,j}=0$ for $j<a$ and $j>b$. In an equivalent but more computationally efficient formulation, one could simply replace the $K$-dimensional $\fx$ in (\ref{QP0}) by the reduced $(b-a+1)$-dimensional counterpart $[f_{X,a},f_{X,{(a+1)}},\cdots,f_{X,b}]^T$ and also replace the $K\times K$ matrix $\C$ by its $K\times(b-a+1)$ counterpart comprised of columns $\{a,a+1,\cdots,b\}$ of $\C$.

\section{Parameter And Regularization Method Selection}\label{sec:selection}
To use the QP method, one must select the number $K$ of histogram bins to represent the empirical density $\hfy$, the regularization parameter $\lambda$, and the regularization method $Q(\fx)$, which we restrict to either Gaussian regularization or second derivative regularization in this paper. For selecting both $\lambda$ and the regularization method, we develop an approach in this section that is based on the Stein's Unbiased Risk Estimate (SURE) method (\cite{stein1981estimation}).

Regarding the choice of $K$, we have found no adverse consequences to using a conservatively large $K$, other than an increase in computational expense. Thus, our recommended approach is to choose a large enough $K$ that it introduces negligible smoothing-related bias in $\hfy$, but not so large that it unnecessarily increases computational expense. Our general rule-of-thumb that we have used in our examples is $K\approx \min\{200,3\sqrt{n}\}$. That is, we select $K$ roughly three times the common $K\approx \sqrt{n}$ rule-of-thumb used in regular histogram density estimation, but no greater than 200. According to \cite{ruppert2002}, selecting a relatively large but fixed number of bins is satisfactory (which can be illustrated by Figure 5 in \cite{QPvignette2018}), and it is not necessary to select the number of bins by some commonly used criteria like GCV or SURE in that using such criteria occasionally causes overfitting.

The focus of Section \ref{sec:SURE} is developing the SURE-like procedure for choosing $\lambda$ and the method of regularization. The SURE-like method is a generalization of Mallows' $C_p$ (\cite{mallows1973some}) criterion that has found widespread use for parameter and model selection in many supervised learning problems (\cite{efron2004estimation}). This method uses an analytical estimate of the expected test error, and, as a result, requires less computational expense than methods like cross-validation. In Section \ref{sec:screeplot}, we develop a graphical method for selecting $\lambda$, which can serve as either a check to avoid using an inappropriate value selected by the SURE-like method (occasionally the automated SURE-like method selects an inappropriate value for $\lambda$) or as a stand-alone method (if an automated method is not needed).

\subsection{A SURE Criterion for Selecting the Regularization Parameter and Method of Regularization}\label{sec:SURE}
Let $\hfy$ denote the histogram (viewed as a $K$-length random vector with the bin locations treated as predetermined) of $Y$ for the ``training" data $\{Y_1,\cdots,Y_n\}$, and let $\hfy^0$ denote the same but for some hypothetical new ``test" sample of $n$ observations of $Y$ drawn from the same distribution but independent of the training data. Let $\hfxl$ denote the estimate of $\fx$ from our QP method described in Section 2 applied to the training data with regularization parameter $\lambda$. We have added a subscript $\lambda$ to $\hfx$ to explicitly indicate its dependence on $\lambda$. To serve as the basis for our approach for selecting $\lambda$, define
\begin{eqnarray}
err&=&\lVert \hfy-\C\hfxl\rVert^2,\mbox{and}  \label{err} \\	
Err&=&\lVert\hfy^0-\C\hfxl\rVert^2	\label{Err}
\end{eqnarray}
as the training and test error, respectively. As in the standard SURE method, our approach is to select the value of $\lambda$ that minimizes the expected test error $\text{SURE}(\lambda) \equiv \E[Err]$, i.e.
\begin{equation}\label{lmd}
\lambda_\text{SURE}= \argmin_{\lambda} \text{SURE}(\lambda) = \argmin_{\lambda} \E[Err],
\end{equation}
where the expectation is with respect to both the training and the test data.

In the remainder of Section \ref{sec:SURE} we derive a tractable approximation for $\text{SURE}(\lambda)$ in Eq.~(\ref{lmd}), and in later sections we demonstrate that it usually provides an effective means to select $\lambda$. In this respect, as a criterion to select $\lambda$, $\text{SURE}(\lambda)$ represents a reasonable balance between tractability and meaningfulness as a measure of quality of the estimate $\hfxl$.

The derivation of $\text{SURE}(\lambda)$ in our context follows the standard derivation used in other SURE-type approaches (\cite{mallows1973some}; \cite{stein1981estimation}; \cite{efron2004estimation}) and begins with the well-known covariance penalty result (see the Appendix or \cite{efron2004estimation}):
\begin{equation}\label{Errexp}
\text{SURE}(\lambda)= \E[Err]=\E[err]+2\mathrm{tr}[\COV(\C\hfxl,\hfy)],
\end{equation}							
where $\COV(\C\hfxl,\hfy)$ denotes the $K\times K$ cross-covariance matrix between the random vectors $\C\hfxl$ and $\hfy$. To estimate $\text{SURE}(\lambda)$, we estimate the two terms in the right-hand-side of (\ref{Errexp}) separately. As an estimate of the expected training error, we use the observed training error in (\ref{err}), which is commonly done in SURE approaches. That is, we use $\widehat{\E}[err]=err=\lVert \hfy-\C\hfxl\rVert^2$.

To estimate the covariance penalty term in Eq.~(\ref{Errexp}), we require a closed-form expression for $\hfxl$. For the general constrained QP formulations discussed in Section \ref{sec:shapeQP}, $\hfxl$ must be solved by numerical optimization, and no closed-form expression for $\hfxl$ exists. However, for a slightly simplified QP formulation with only equality constraints,
\begin{eqnarray}
\hfxl&=&\argmin_{\fx} \lVert \hfy-\C\fx\rVert^2+\lambda Q(\fx) \nonumber\\
s.t.& &\delta\1^T \fx=1,	\label{QP00}	
\end{eqnarray}								
we can apply the Lagrange multiplier method to find a closed-form solution of the form $\hfxl=\B\hfy+\bb$, where $\B$ and $\bb$ are functions of $\lambda$.  Via Lagrange multipliers, it is not difficult to derive that $\B$ and $\bb$  in the solution to (\ref{QP00}) are
\begin{eqnarray*}
	\B&=&\left\{\D^{-1}-\frac{(\D^{-1}\1\1^T\D^{-1})}{\1^T\D^{-1}\1}\right\}\C^T\\
	\bb&=&\left\{\D^{-1}-\frac{(\D^{-1}\1\1^T\D^{-1})}{\1^T\D^{-1}\1}\right\}\bm{\mathrm{q}}+\frac{\D^{-1} \1}{\delta\1^T \D^{-1}\1}.
\end{eqnarray*}
In the preceding expressions, for second derivative regularization [i.e., for $Q(\fx)=\lVert \D_2 \fx\rVert^2$], $\D=(\C^T\C+\lambda\D_2^T \D_2)$, and $\bm{\mathrm{q}}=\0$ is a $K$-length vector of zeros. For Gaussian regularization [i.e., for $Q(\fx)=\lVert \fx-\freg\rVert^2$], $\D=(\C^T\C+\lambda\bm{\mathrm{I}})$, and $\bm{\mathrm{q}}=\freg$.

Using the preceding, the covariance penalty term in Eq.~(\ref{Errexp}) can be simplified to
\begin{equation}\label{Covexp}
\COV(\C\hfxl,\hfy)=\COV[\C(\B\hfy+\bb),\hfy]=\C\B\bS,
\end{equation} 						
where $\bS=\COV(\hfy)$ denotes the $K\times K$ covariance matrix of $\hfy$. To obtain an expression for $\bS$, note that in our QP formulation, the density $\hfy$ is taken to be heights of the histogram bins (scaled to represent a density). Thus, the vector $n\delta\hfy$ follows a multinomial distribution
\[
n\delta\hfy\sim \mathrm{MN}(\bm{\mathrm{p}},n),
\]
where $\bm{\mathrm{p}}=\delta\fy=\delta\C\bm{\mathrm{f}}_{X,\lambda}$ is a $K$-dimensional vector with elements denoted by $(p_i)_{i=1}^K$.  From the properties of a multinomial distribution, we know that $\E[n\delta\hfy]=n\bm{\mathrm{p}}$, $\VAR[n\delta\hfyi]=np_i(1-p_i)$ and $\COV[n\delta\hfyi,n\delta\hfyj]=-np_ip_j$ for $i\neq j$. Thus, given that $\COV[n\delta\hfy]=n^2\delta^2\bS$, the elements of $\bS$ are $\bS_{ii}=p_i(1-p_i)/n\delta^2$ $(i=1,\cdots,K)$ and $\bS_{ij}=-p_ip_j/n\delta^2$ $(i\neq j)$, i.e.,
\begin{equation}\label{Sexp}
\bS=\frac{\mathrm{diag}\{p_i \}_{i=1}^K-\bm{\mathrm{p}}\bm{\mathrm{p}}^T}{n\delta^2}=\frac{\mathrm{diag}\{\delta f_{Y,i}\}_{i=1}^K-\delta^2 \fy\fy^T}{n\delta^2}.
\end{equation}

An estimate of $\bS$ is obtained by replacing the true $\fy$ in (\ref{Sexp}) with the observed histogram $\hfy=(\widehat{f}_{Y,i})_{i=1}^K$. Combining this with Eqs.(\ref{Errexp}) and (\ref{Sexp}), the estimate of $\text{SURE}(\lambda)$ becomes
\begin{equation}\label{ExpErr}
\widehat{\text{SURE}}(\lambda)= \lVert \hfy-\C\hfxl\rVert^2+2\mathrm{tr}\left[\C\B\frac{\mathrm{diag}\{\delta \widehat{f}_{Y,i}\}_{i=1}^K-\delta^2\hfy\hfy^T}{n\delta^2}\right],				
\end{equation}
which, for small $\delta\hfy$ (i.e., small multinomial probabilities $\bm{\mathrm{p}}$, which will generally be the case if one chooses an appropriate number $K$ of histogram bins), can be approximated by
\begin{equation}\label{hExpErr}
\widehat{\text{SURE}}(\lambda)\approx\lVert \hfy-\C\hfxl\rVert^2+2\mathrm{tr}\left[\C\B\frac{\mathrm{diag}\{\widehat{f}_{Y,i}\}_{i=1}^K}{n\delta}\right].	\end{equation}

In the right hand side of Eq.~(\ref{hExpErr}), $\hfxl$ and $\B$ depend on $\lambda$, and the other terms do not. To simplify notation, rewrite Eq.~(\ref{hExpErr}) as $\widehat{\text{SURE}}(\lambda)\approx\lVert\hfy-\C\hfxl\rVert^2+g(\lambda)$, where $g(\lambda)=2\mathrm{tr}\left[\C\B\mathrm{diag}\{\widehat{f}_{Y,i}\}_{i=1}^K\right]/(n\delta)$. Our SURE-like criterion (\ref{lmd}) for selecting the optimal regularization parameter with only the integrate-to-one constraint becomes:
\begin{eqnarray}\label{lmdSURE1}
\lmds&=&\argmin_{\lambda}⁡\lVert \hfy-\C\hfxl\rVert^2+g(\lambda)\nonumber\\
\text{where}& & \hfxl=\argmin_{\fx}  \lVert\hfy-\C\fx\rVert^2+\lambda Q(\fx)\nonumber\\
& &  s.t.\quad\delta\1^T \fx=1.	
\end{eqnarray}								

The preceding SURE derivation is not strictly valid if more constraints than the integrate-to-one constraint are used. With additional constraints, one might consider using (\ref{lmdSURE1}) to select the best $\lambda$, and then reconducting the optimization (\ref{QP}) for that value of $\lambda$ with all constraints included to produce the final estimate $\hfx$. However, we have found the following more compact modification of (\ref{lmdSURE1}) to be generally more effective. Namely, our SURE-like approach for selecting the optimal regularization parameter with multiple constraints is:
\begin{eqnarray}\label{lmdSURE2}
\lmds&=&\argmin_{\lambda}⁡\lVert \hfy-\C\hfxl\rVert^2+g(\lambda)\nonumber\\
\text{where}&& \hfxl=\argmin_{\fx}  \lVert\hfy-\C\fx\rVert^2+\lambda Q(\fx)\nonumber\\
&&  s.t.\quad\text{all relevant constraints are satisfied.}	
\end{eqnarray}

The method of regularization (Gaussian or second derivative) can also be selected via SURE. This is accomplished by performing the optimization in (\ref{lmdSURE2}) separately, for both regularization methods, and then choosing the method that gives the smallest SURE expected error $\lVert \hfy-\C\hfxl\rVert^2+g(\lambda)$. 	

\subsection{A Graphical Scree-plot Approach for Selecting the Regularization Parameter}\label{sec:screeplot}
\begin{figure}[!tbp]
	\centering
	\subcaptionbox{\label{fig:fig3a}}{\includegraphics[width=1\textwidth]{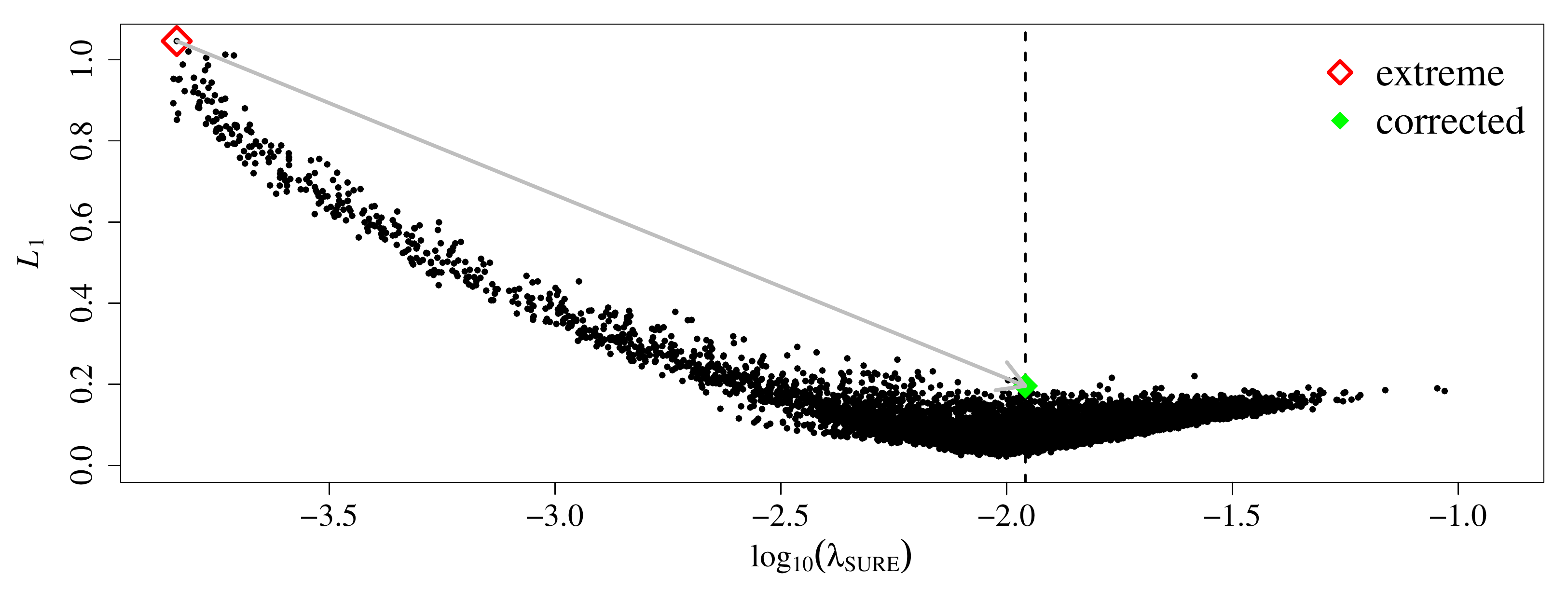}}
	\hfill
	\subcaptionbox{\label{fig:fig3b}}{\includegraphics[width=0.49\textwidth]{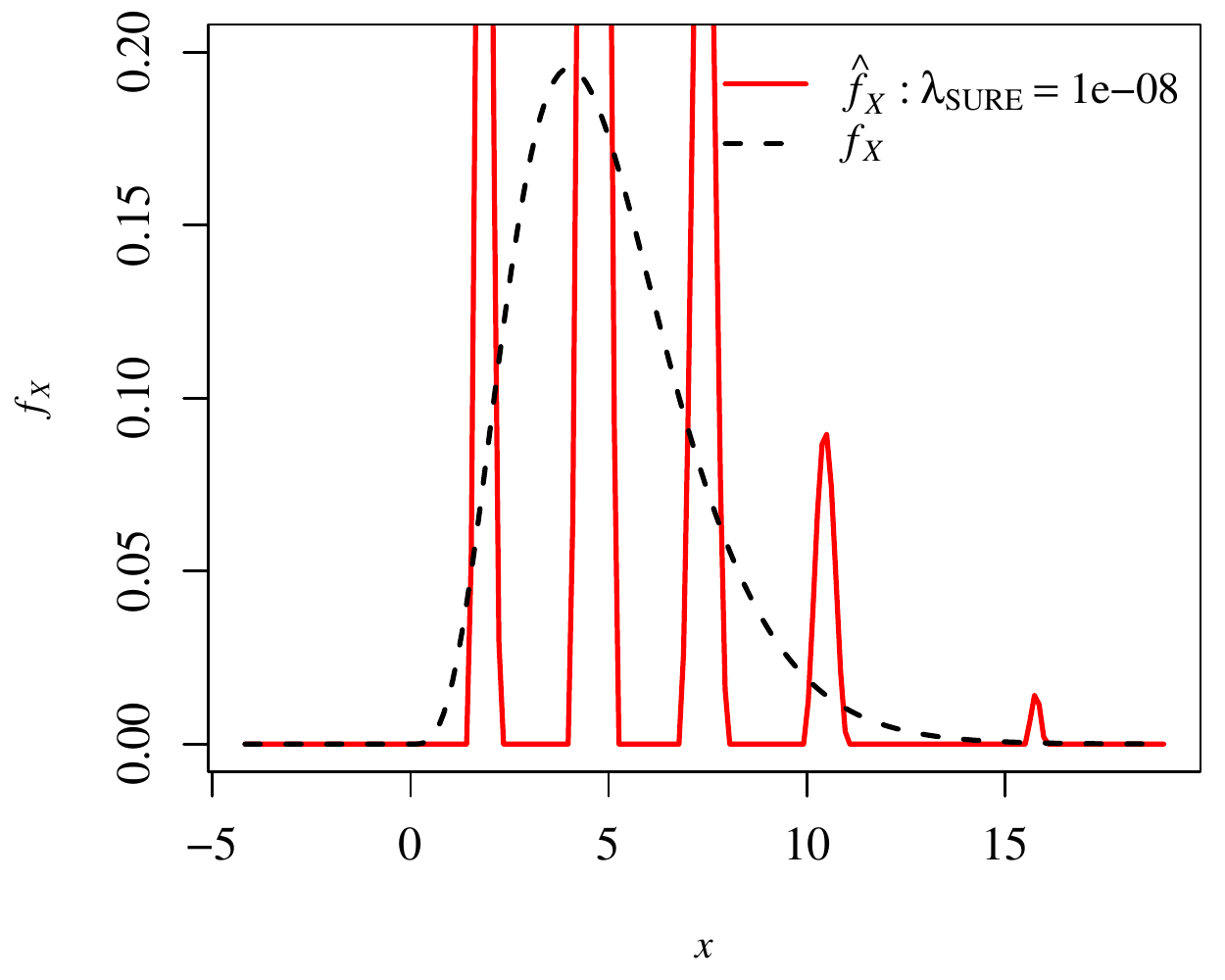}}
	\hfill
	\subcaptionbox{\label{fig:fig3c}}{\includegraphics[width=0.49\textwidth]{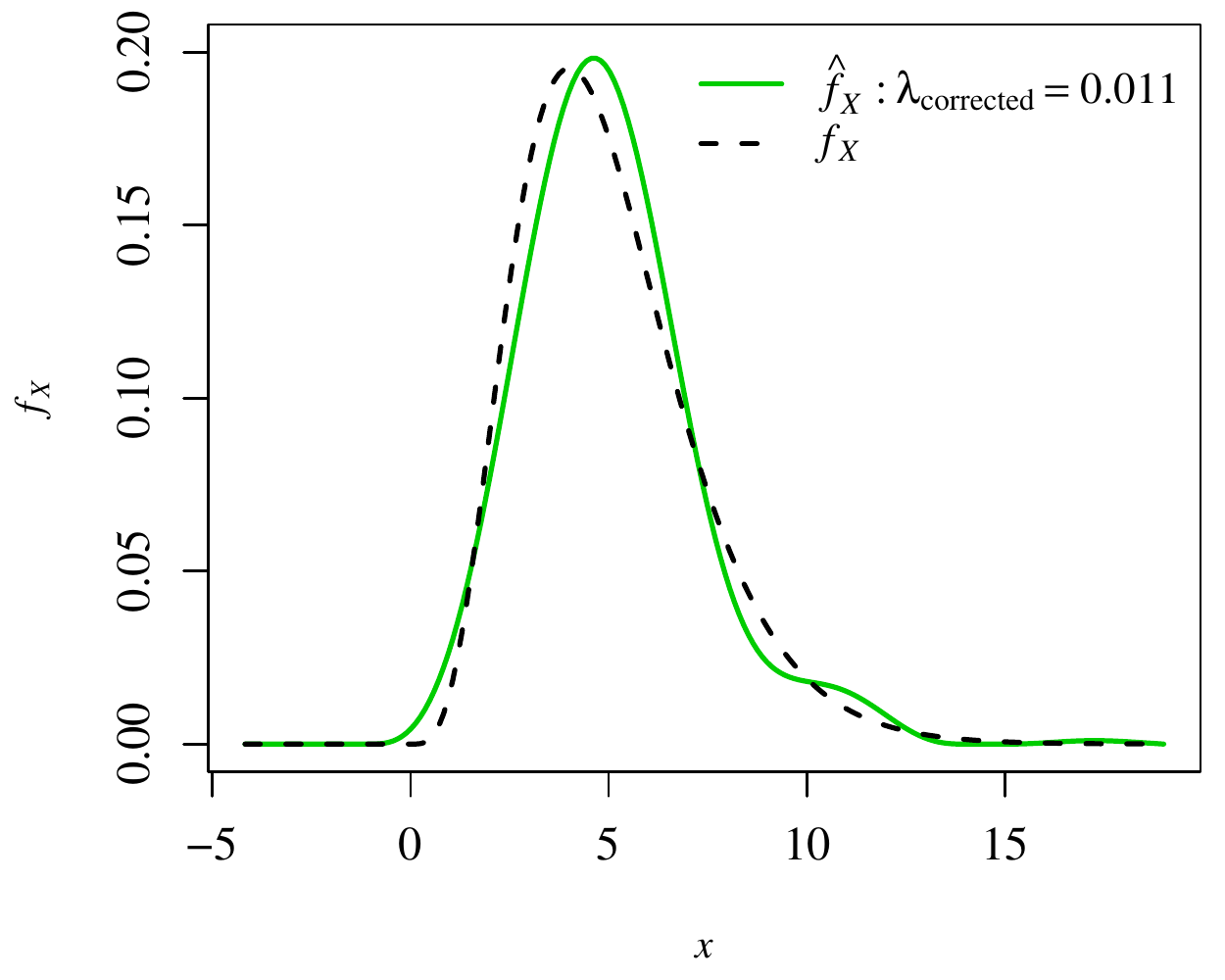}}
	\caption{(a) Scatter plot of the $L_1$ error measure versus $\log_{10}(\lmds)$ for 8,000 replicates of the $Gamma(5,1)$ example; the dashed vertical line corresponds to $\lambda=0.011$. (b) A poor pdf estimate for the replicate corresponding to the open red diamond in the upper-left corner of panel (a), for which the estimated $\lmds$ is much too small.  (c) A much-improved estimate $\hfx$ for the data for the same replicate featured in panel (b), but using a corrected $\lambda= 0.011$ obtained from the scree plot method. The $L_1$ error of this improved estimate is represented by the solid green diamond in panel (a).\label{fig:L1dist}}
\end{figure}

On a relatively small percentage of Monte Carlo (MC) replicates in the numerous examples that we have investigated, $\lmds$ from (\ref{lmdSURE2}) is chosen inappropriately. This is usually because $\lmds$ is chosen too small, and the QP method results in a high-variance estimator. This is illustrated in Fig.~\ref{fig:L1dist} for 8,000 replicates of the same $Gamma(5,1)$ example considered in Fig.~\ref{fig:introplot}. For each replicate, we generated a random sample of size $n=5,000$ for the random variables $X$ and $Z$ and then used the $5,000$ values of $Y = X + Z$ as the observed data. For all replicates in Fig.~\ref{fig:L1dist}, $\hfx$ was estimated using the QP method with only the two universal shape constraints of integrate-to-one and nonnegativity. Fig.~\ref{fig:fig3a} plots the estimation error measure $L_1(\widehat{f}_X,f_X)=\int|\widehat{f}_X(x)-f_X(x)|\mathrm{d}x$ against $\log_{10}(\lmds)$. We observe that 9.4\% of replications have $L_1$ error more than twice the median $L_1$ error (the median is 0.089), and about 5.7\% have error more than three times the median. The QP estimator in Fig.~\ref{fig:fig3b} corresponds to one of the occasional replicates for which $\lmds$ is extremely underestimated, and its $L_1$ error is represented by the open red diamond in Fig.~\ref{fig:fig3a}. In comparison, Fig.~\ref{fig:fig3c} shows a much-improved estimation result using a corrected $\lambda=0.011$ (corrected via the scree plot method, described below) for the data from the same replicate featured in Fig.~\ref{fig:fig3b}, and the $L_1$ error of the improved result is represented by the solid green diamond in Fig.~\ref{fig:fig3a}. Notice that the $L_1$ error is reduced from $1.05$ to $0.19$, a level that is far below the level for the extreme case and much more consistent with typical cases (twice the median $L_1$ error). 

\begin{figure}[!tbp]
	\centering
	\includegraphics[width=0.55\textwidth]{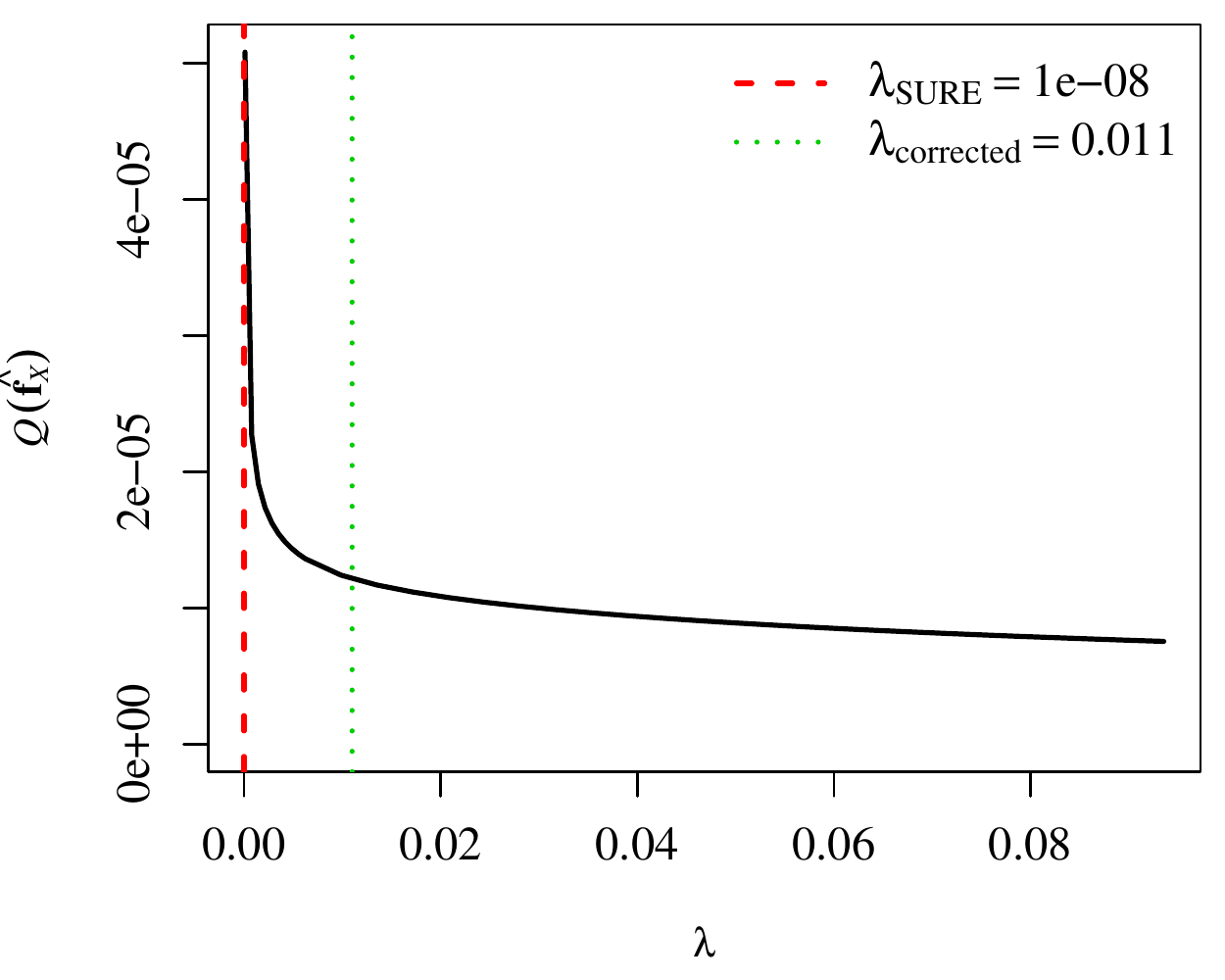}
	\caption{Scree plot for the replicate featured in Fig.~\ref{fig:fig3b}. The vertical red dashed line indicates the value for $\lmds$, which was much too small on this example and resulted in the poor pdf estimate in Fig.~\ref{fig:fig3b}. The vertical green dotted line indicates the corrected $\lambda$, chosen to the right of the elbow, which resulted in the substantially better pdf estimate shown in Fig.~\ref{fig:fig3c}.}\label{fig:screeplot}
\end{figure}

The corrected $\lambda = 0.011$ was obtained by inspection of the scree plot in Fig.~\ref{fig:screeplot}, which is a simple graphical method that we have found to provide an effective means for selecting an appropriate regularization parameter and avoiding the poor pdf estimation that results on the occasional replicates in which the SURE-like method selects an inappropriate value for $\lambda$. 
As illustrated in Fig.~\ref{fig:screeplot}, the scree plot is a plot of $Q(\hfx)$ versus $\lambda$, and we look for the elbow in the plot. Namely, our scree-plot choice for $\lambda$ is the smallest value of $\lambda$ that is comfortably to the right of the elbow. This is analogous to how a plot of the norm of the estimated ridge regression coefficient vector versus the regularization parameter is used to select the regularization parameter in ridge regression (\cite{hoerl1970}).

A number of conclusions can be drawn from Fig.~\ref{fig:fig3a}. First, we note that the best single value for $\lambda$ in this $Gamma(5,1)$ example was roughly $\lambda= 0.011$, which we found by comparing the MC average $L_1$ error values for a range of fixed $\lambda$ values (the results of which are omitted, for brevity). We refer to this best single value of $\lambda$ as the ``oracle" value. The oracle value $\lambda= 0.011$ is also somewhat apparent from Fig.~\ref{fig:fig3a}, because if we smooth the scatter plot, the smoothed $L_1$ error would be smallest at approximately $\lambda= 0.011$. Also from Fig.~\ref{fig:fig3a}, the mode of the 8,000 $\lmds$ values produced over the 8,000 MC replicates was also $0.011$, the same as the oracle value, and in this respect the SURE-like method did an overall good job of selecting $\lambda$. 

Another conclusion from Fig.~\ref{fig:fig3a} is that on replicates for which the SURE-like method did a poor job of selecting $\lambda$, resulting in large $L_1$ error, it was always because $\lmds$ was \textit{underestimated}. Moreover, and significantly, for all of the replicates with $\lmds$ underestimated, the scree plots (not shown here, for brevity) always looked very much like the one shown in Fig.~\ref{fig:screeplot}, and the corrected $\lambda$ (selected to the right of the elbow) always substantially improved the pdf estimate, as in Figure~\ref{fig:fig3c}. We conclude that the scree plot provides a simple and effective means of selecting $\lambda$. 



We now illustrate the performance of the automated SURE-like $\lambda$ selection method and also how to select $\lambda$ (or correct an underestimated $\lmds$) using the scree plot with one replicate of the simple example, i.e $n=5000$, $X\sim Gamma(5,1)$, $Z\sim N(0,\sigma_Z^2=3.2)$, and $Y=X+Z$.

\begin{figure}[h]
	\centering
	\includegraphics[width=0.6\textwidth]{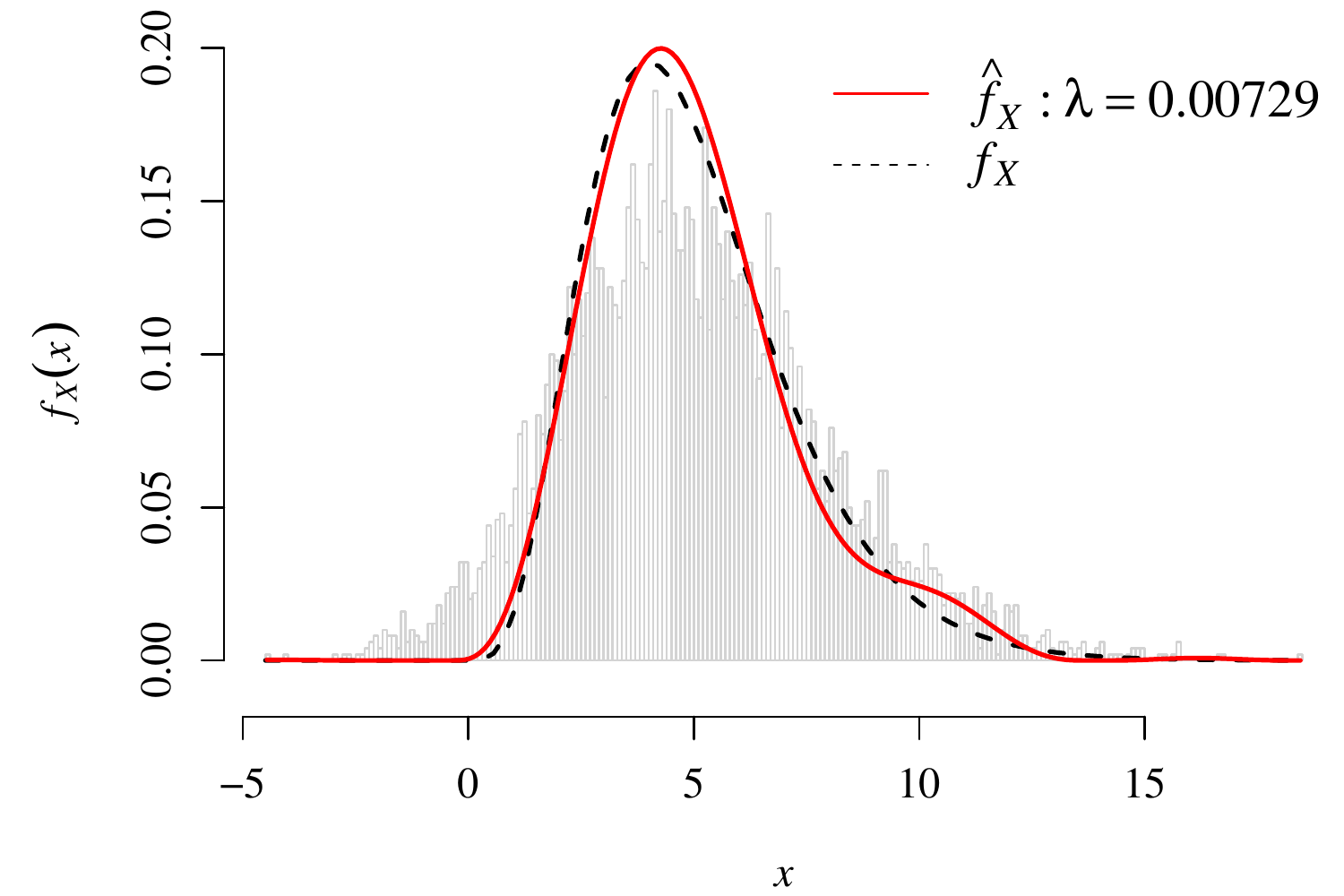}
	\caption{Histogram of the observed data $Y$ together with the true pdf of $X$ (black dashed curve) and the QP pdf estimator (red solid curve) using $\lmds=0.0073$ function.}\label{fig:qpsure}
\end{figure}

Fig.~\ref{fig:qpsure} shows the histogram for the sample of observations of $Y$ as well as the true and estimated pdf of $f_X$ (black dashed curve and red solid curve, respectively), and the estimated pdf uses only the nonnegativity and integrate-to-one constraints. The automatically selected regularization parameter for this example is $\lmds=0.007295$, and we can see this automated selection worked quite well for this example despite $\hfx$ having a slight oscillation on the right tail. 

\begin{figure}[!tbp]
	\centering
	\subcaptionbox{\label{fig:Rscreeplot}}{\includegraphics[width=0.49\textwidth]{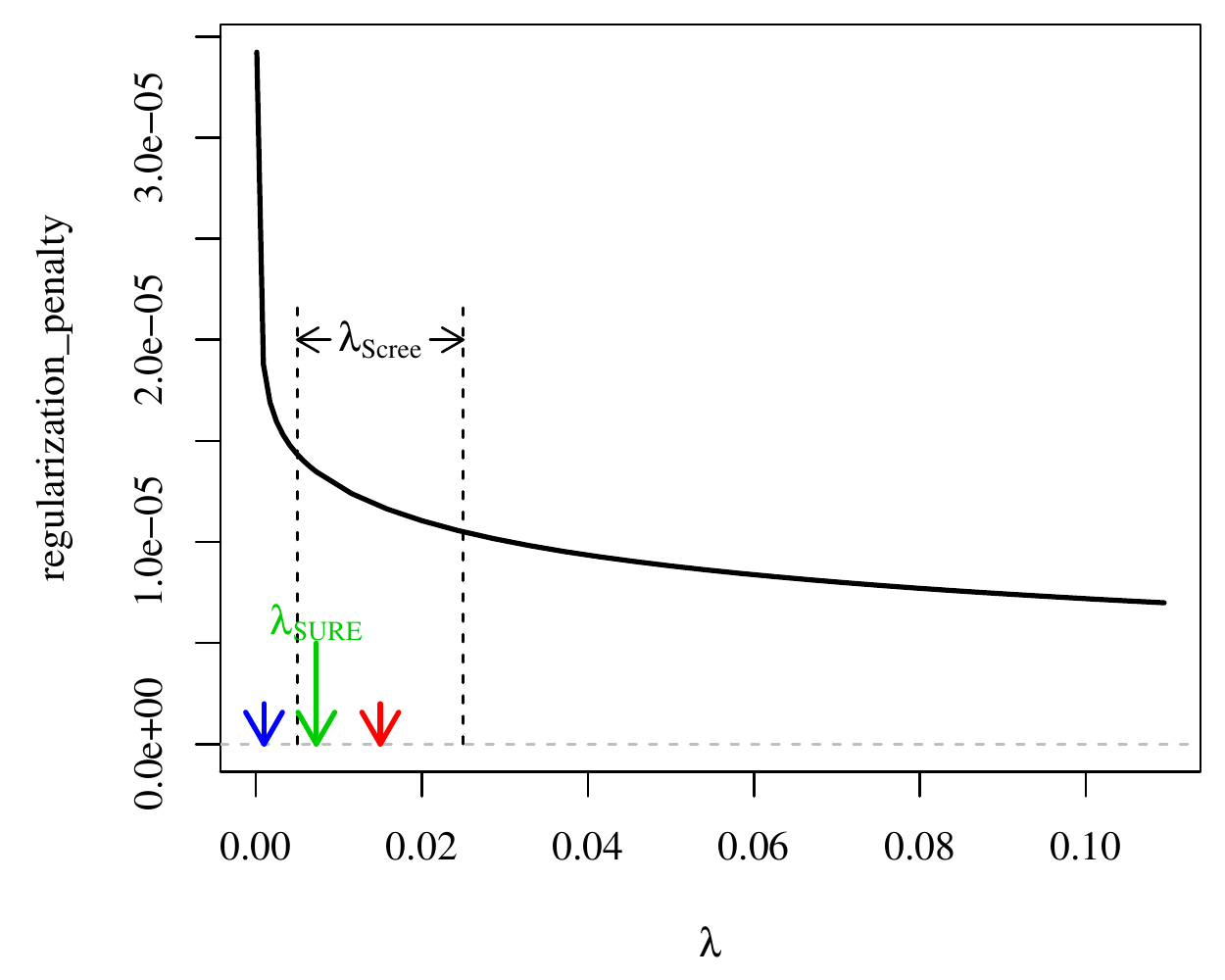}}
	\hfill
	\subcaptionbox{\label{fig:screeqp}}{\includegraphics[width=0.49\textwidth]{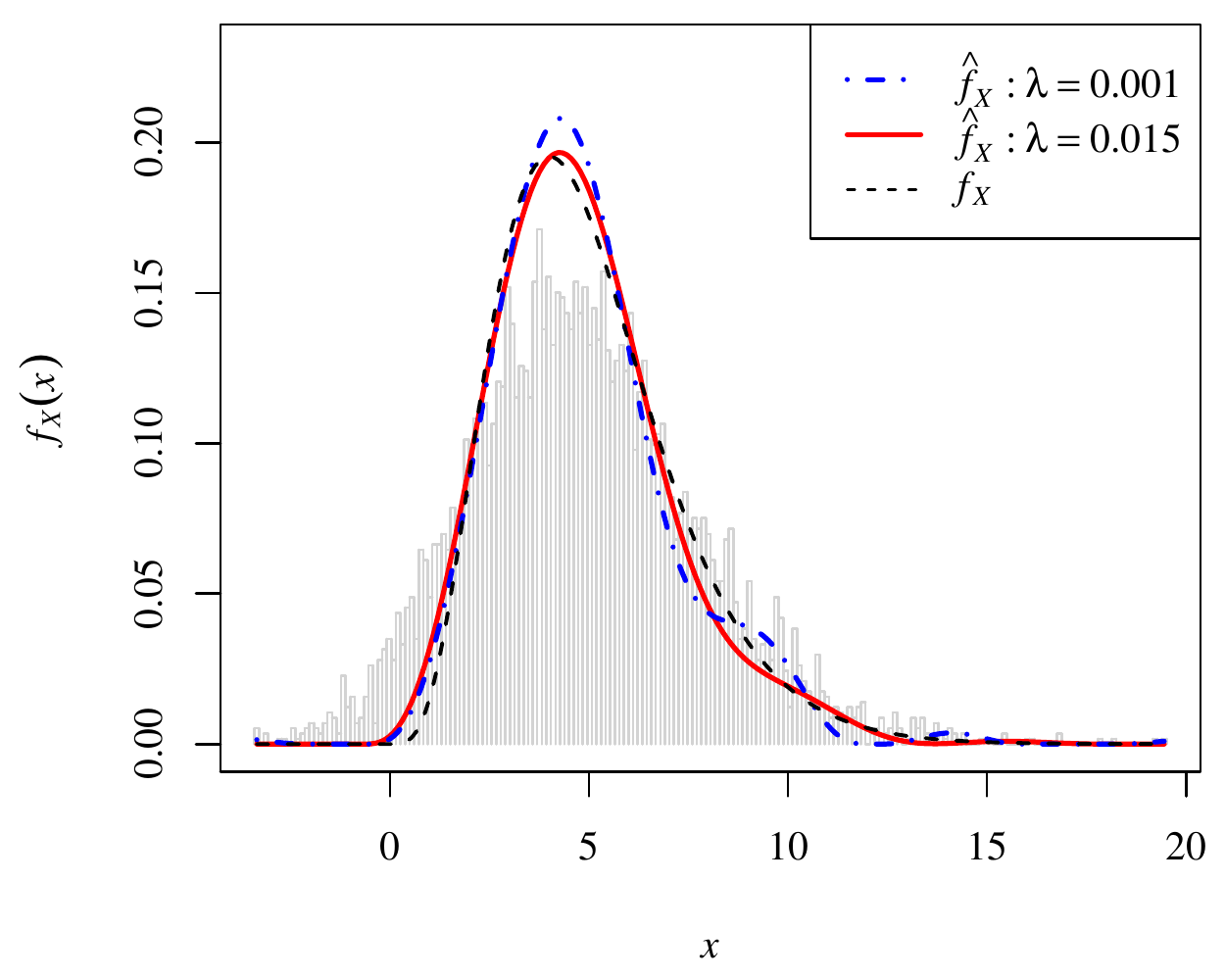}}
	\caption{(a): Scree plot for the data depicted in Fig.~\ref{fig:qpsure}. The long green arrow indicates the value of $\lmds$ used in Fig.~\ref{fig:qpsure}, and the two dashed vertical lines roughly indicate the range of $\lambda$ values suggested by the scree-plot method. (b): The histogram of the observed data $Y$ together with the QP estimators for the ``inappropriately small" $\lambda$ value (0.001) indicated by the blue arrow in panel (a) and for an appropriate $\lambda$ value (0.015) indicated by the red arrow in panel (a), which falls within the $\lmdsr$ range.}\label{fig:qpscree}
\end{figure}

As an alternative to using the SURE-like method to select $\lambda$, or as a check that $\lmds$ is appropriate, we can use the scree-plot method. The scree plot is constructed by repeatedly calculating the QP estimator $\hfx$ for a set of values of $\lambda$, and the scree plot for the data depicted in Fig.~\ref{fig:qpsure} is shown in Fig.~\ref{fig:Rscreeplot}. The long green arrow in Fig.~\ref{fig:Rscreeplot} indicates $\lmds= 0.0073$, which was obtained from the automated SURE-like method. We have added the two dashed vertical lines to indicate roughly what may be viewed as the lower and upper bounds of the candidate $\lambda$ values suggested by the scree-plot method, and we denote any $\lambda$ falling in this range as $\lmdsr$. The arrow to the left of $\lmds$ indicates a value ($\lambda= 0.001$) that falls substantially below the $\lmdsr$ range and is clearly to the left of the elbow. The arrow to the right of $\lmds$ indicates a value ($\lambda= 0.015$) that is within the $\lmdsr$ range. The QP pdf estimators corresponding to these two $\lambda$ values are shown in Fig.~\ref{fig:screeqp}, from which we can see that using a $\lambda$ value that is too small results in apparent tail oscillations and over-estimation in the middle quantiles, whereas using a moderate size of $\lambda$ within the range of $\lmdsr$ smooths out the oscillations of the QP estimator without deteriorating (oversmoothing) its performance in the middle quantiles. For this particular replicate, the SURE-like method provides a regularization parameter $\lmds$ that falls within the range of $\lmdsr$ and results in good performance. However, as discussed earlier, there are replicates on which $\lmds$ is chosen too small, and when this happens, the scree plot clearly indicates this (because $\lmds$ falls to the left of the elbow, as in Fig.~\ref{fig:screeplot}), so that a more appropriate $\lambda$ can be selected to improve the performance of the QP method. 

\section{Discussion and Performance Comparisons}\label{sec:discussion}
To further investigate the performance of our QP method and compare it to the other two density deconvolution methods KD and PC, we investigate two examples using the Monte Carlo simulations. The first is $X\sim Exp(0.447)$ (mean$=2.24$ and variance$=5$) and $Z\sim N(0,\sigma_Z^2=3.2)$; and the second is $X\sim Gamma(5,1)$ (mean$=5$ and variance$=5$), and $Z\sim N(0,\sigma_Z^2=3.2)$. We choose a Gaussian distribution for the noise not only because Gaussian noise is common, but also because it is considered supersmooth and notoriously difficult to deconvolve. Moreover, the smaller the signal-to-noise ratio $\sigma_X^2/\sigma_Z^2$, the more difficult is the deconvolution. For both examples, the signal-to-noise ratio is about 1.56, which is relatively small. Consequently, we use a relatively large sample size of $n=5000$ for all numerical studies. We use various sets of constraints, which we denote by the following initials:  integrate-to-one ($i$), nonnegativity ($n$), tail monotonicity ($m$), tail convexity ($c$), unimodality ($u$) and support ($s$). We denote combinations of constraints by combinations of the initials. For convenient reference, we display the constraints we consider and their acronyms in Table~\ref{tab:acronyms}. For all examples, the basic constraints $in$ are used for the QP approach because they apply to all pdfs. In order to have a more common basis for comparison, the KD and PC estimators are scaled retrospectively so that they integrate-to-one and are nonnegative, which we refer to as retro-$in$. This also improved their performances overall, relative to not using the retro-$in$ adjustment. More specifically, in the retro-$in$ approach, all negative elements of $\hfx$ are reset to zero, after which all elements of $\hfx$ are scaled proportionately so that $\hfx$ integrates-to-one. 

\begin{table}[!h]
    \centering
    \caption{Acronyms for the shape constraints}
    \begin{tabular}{cccccc}
        \hline
       integrate-to-one& nonnegativity& tail monotonicity& tail convexity& unimodality & support  \\
        \hline
       $i$ & $n$ & $m$ & $c$ & $u$& $s$\\
        \hline
    \end{tabular}
    \label{tab:acronyms}
\end{table}

In Section \ref{sec:discuss_in}, we compare the quantile estimation performances of the QP method with only $in$ constraints versus the KD and PC methods with retro-$in$, and in Section \ref{sec:discuss_shape} we investigate the improvement in QP estimation performance that results from including shape constraints. Section \ref{sec:discuss_pdf} compares performances directly in terms of pdf estimation.  

\subsection{Performance Comparisons with Only $in$ Constraints}\label{sec:discuss_in}
To distinguish estimation performance in left tail, right tail and central portions of the distribution, we consider the median absolute errors ($MAE$s) for estimating the nine quantiles corresponding to probabilities $p\in \{0.01, 0.05, 0.1, 0.25, 0.5, 0.75, 0.9, 0.95, 0.99\}$ for each example. We use $MAE$ rather than mean square error, because the former is more robust to occasional outliers that can occur when $\lambda$ is underestimated by the SURE method. Downweighting such outliers is justified, because our graphical scree-plot method for choosing $\lambda$ can effectively eliminate most of these outliers (see, e.g., Fig. \ref{fig:fig3a}). We use the automated $\lmds$ when selecting $\lambda$ in the MC analysis because the scree-plot method for selecting $\lambda$ requires user input based on inspection of the scree plots. 

Recall that the KD estimator (\cite{carroll1988optimal}) is:
\[
\widehat{f}_X(x)=\frac{1}{2\pi n}\sum_{j=1}^n\int_{-\infty}^{\infty}\phi_K(h\omega)
\phi_Z^{-1}(\omega) \mathrm{exp}⁡\{-i\omega(x-Y_j)\}\mathrm{d}\omega
\]

We consider two common kernels for KD in the subsequent comparisons. One is the rectangular kernel, $\phi_K(\omega)=I_{[-1,1]}(\omega)$, and the other is the triweight kernel, $\phi_K(\omega)=(1-\omega^2 )^3I_{[-1,1]}(\omega)$. Regarding the selection of the KD bandwidth $h$, we follow the recommendations in \cite{delaigle2004practical} and use the bootstrap method with their rule-of-thumb initial guess $h=\sqrt{2}\sigma_Z \sqrt{\mathrm{log}(n)}$. Henceforth, we will refer to QP's regularization parameter $\lambda$ and KD's bandwidth parameter $h$ both as ``regularization parameters". We will compare the performances of five estimators, and they are QP with only the $in$ constraints (denoted as QP\textsubscript{$in$}), QP with additional constraints, KD with either rectangular kernel (KD\textsubscript{$rect$}) or triweight kernel (KD\textsubscript{$triw$}), as well as the PC estimator.  

\begin{figure}[!tbp]
	\centering
	\subcaptionbox{$MAE$ ratio corresponding to Table~\ref{tab:expMAE}\label{fig:MAEratio2}}{\includegraphics[width=0.49\textwidth]{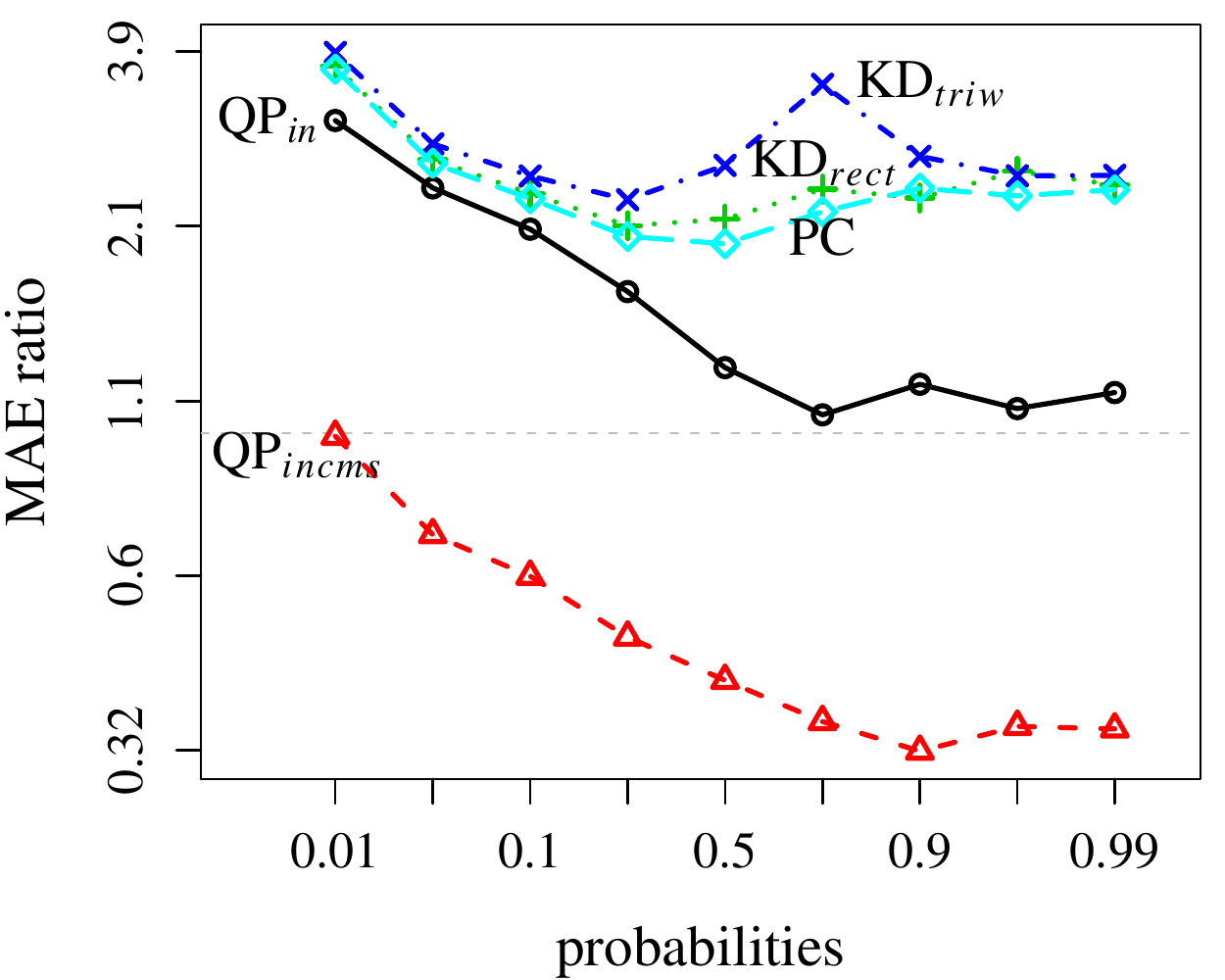}}
	\hfill
	\subcaptionbox{$MAE$ ratio corresponding to Table~\ref{tab:gammaMAE}\label{fig:MAEratio1}}{\includegraphics[width=0.49\textwidth]{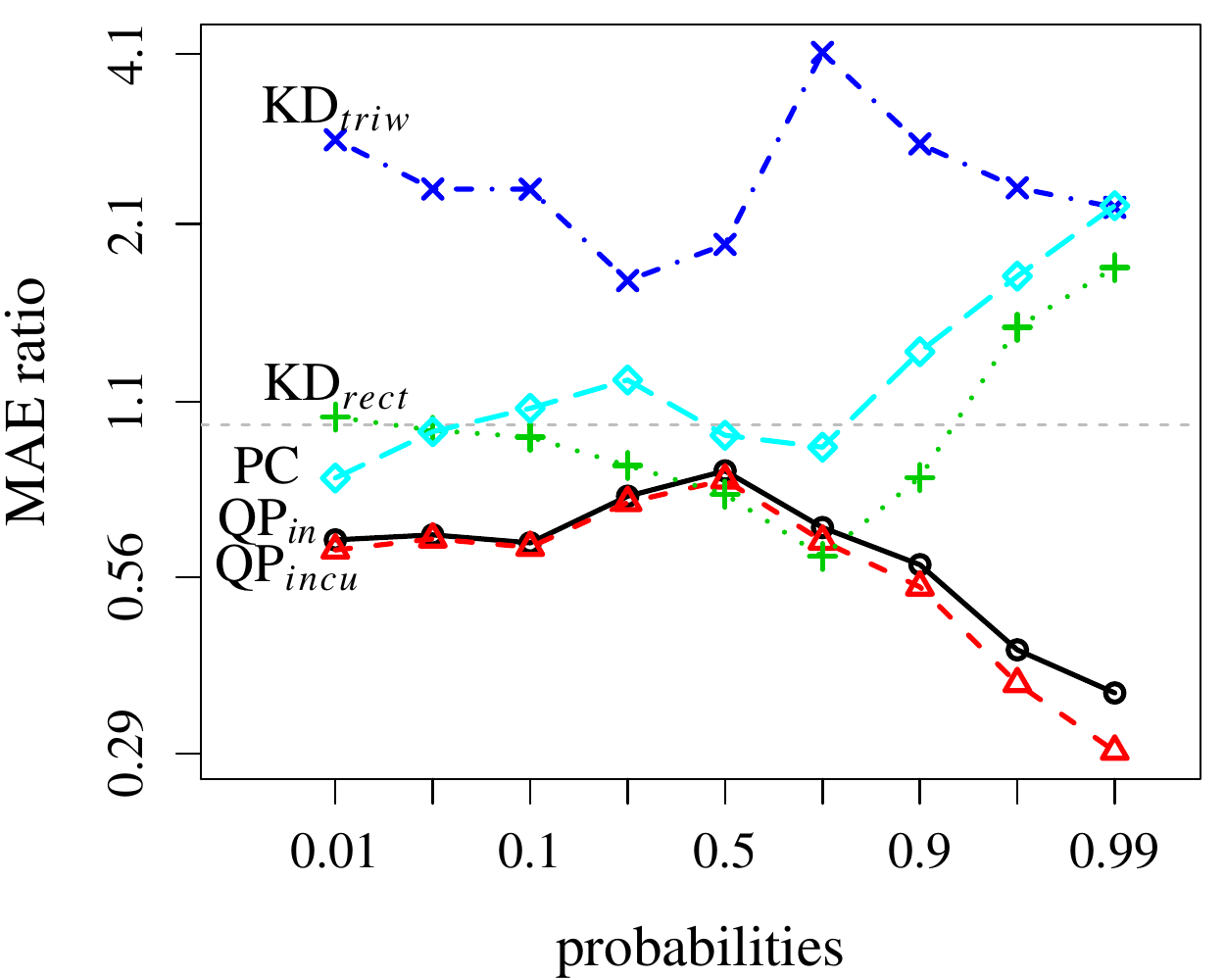}}
	\hfill
	\subcaptionbox{$MAE$ ratio corresponding to Table~\ref{tab:expMAEoracle}\label{fig:MAEratio4}}{\includegraphics[width=0.49\textwidth]{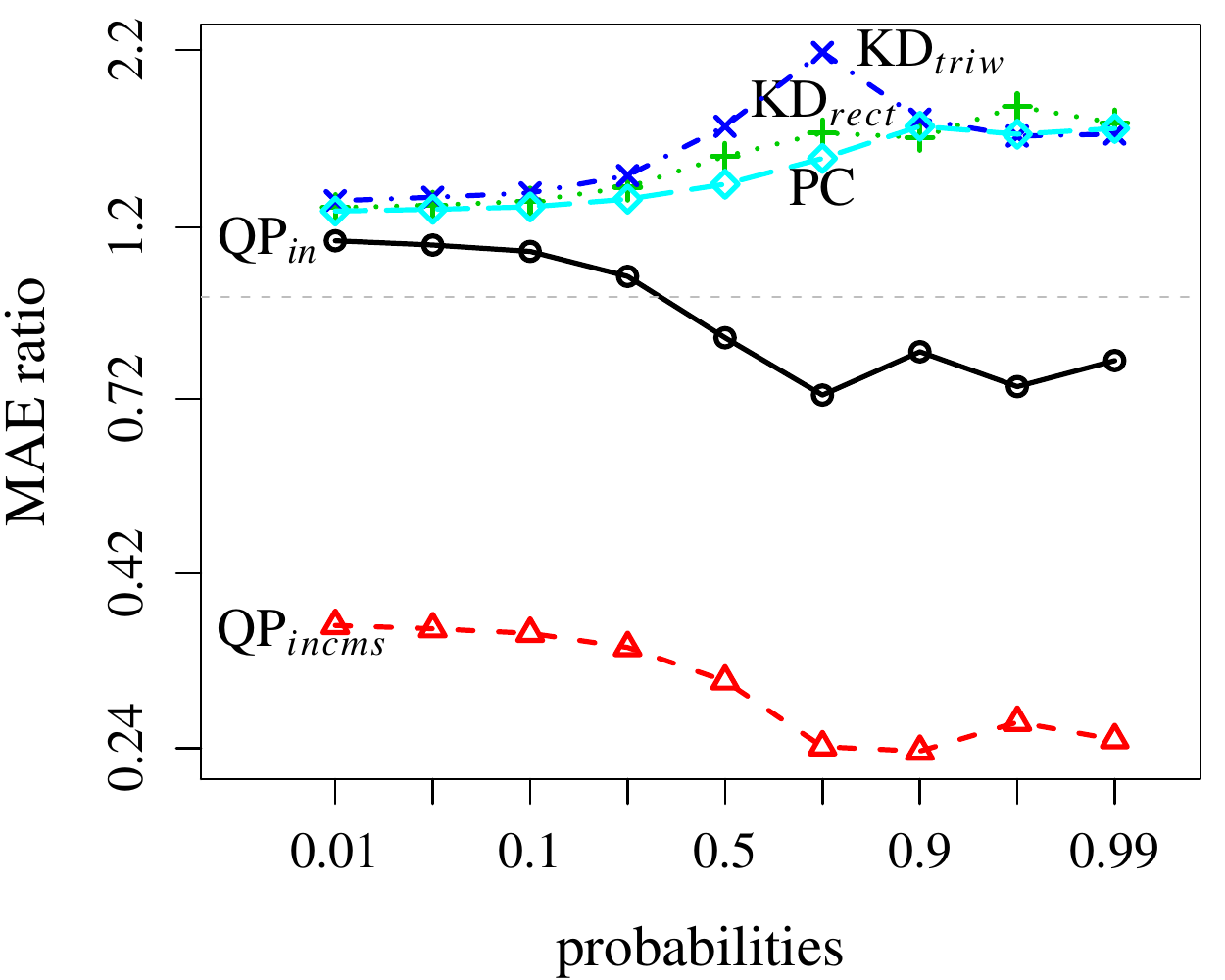}}
	\hfill
	\subcaptionbox{$MAE$ ratio corresponding to Table~\ref{tab:gammaMAEoracle}\label{fig:MAEratio3}}{\includegraphics[width=0.49\textwidth]{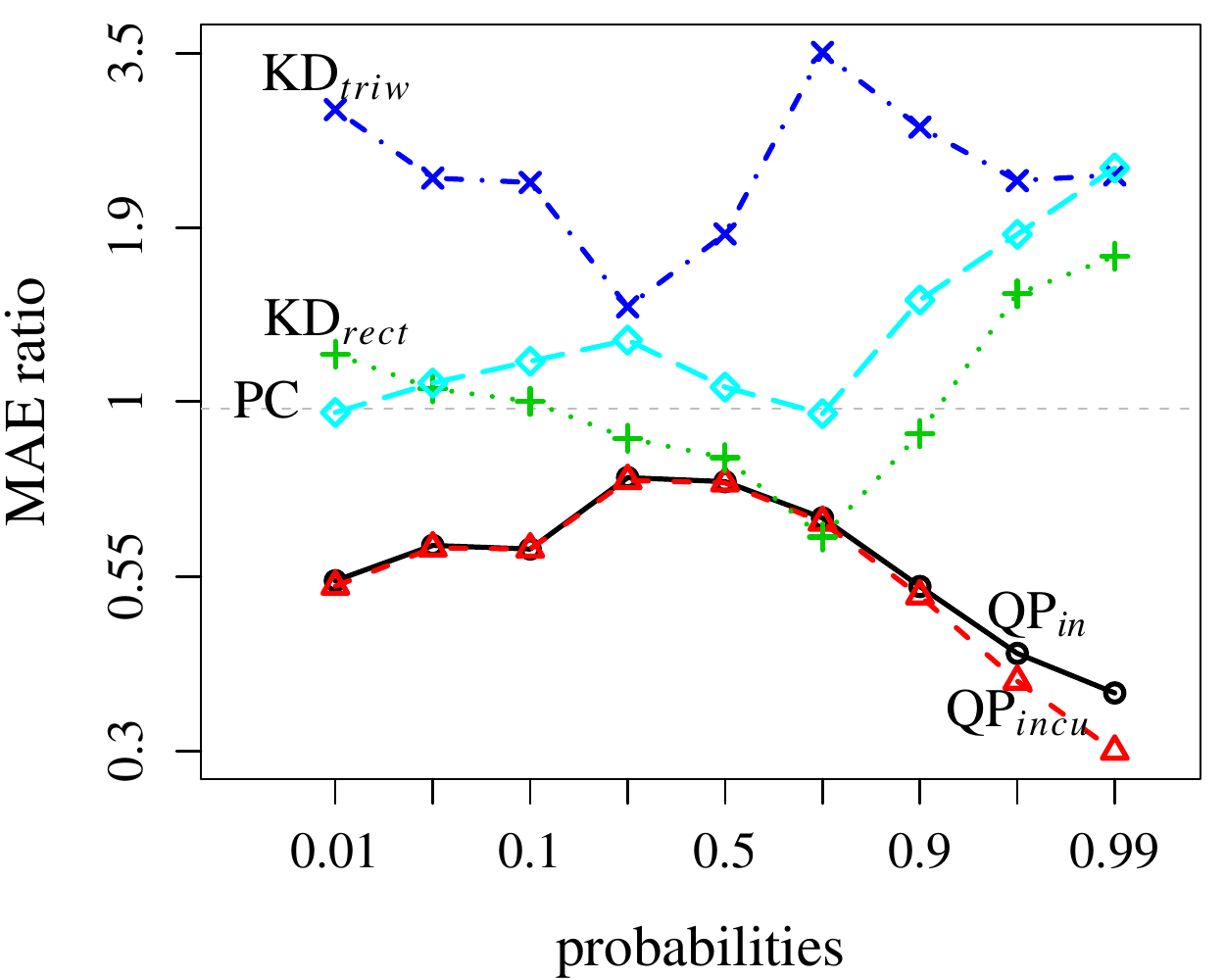}}
	\caption{$MAE$ ratios (plotted in log-scale) corresponding to Tables~\ref{tab:expMAE}--\ref{tab:gammaMAEoracle}. The numerator of the $MAE$ ratio is the $MAE$ for each estimator, and the denominator is the geometric average of the $MAE$ across all five estimators shown in the corresponding tables.}\label{fig:MAEratio}
\end{figure}

Table \ref{tab:expMAE} and Table \ref{tab:gammaMAE} show the quantile $MAE$ results (averaged across all MC replicates) with automated selection of the regularization parameters for the exponential and gamma examples, respectively. 
For the $MAE$s, the median was obtained over 500 and 8,000 MC replicates for the exponential and gamma examples, respectively. We choose the number of replicates so that the standard errors of the $MAE$s are less than or equal to 1\% of the $MAE$ value. Figures~\ref{fig:MAEratio2} and ~\ref{fig:MAEratio1} provide a visual display of the relative performances of all five estimators in Tables~\ref{tab:expMAE} and~\ref{tab:gammaMAE}, respectively. Each curve is the $MAE$ ratio (plotted in log-scale) across all nine probabilities for a particular estimator. The numerator of the $MAE$ ratio is the $MAE$ of the corresponding estimator, and the denominator is the geometric average of the $MAE$s across all five estimators.

From Tables \ref{tab:expMAE} and~\ref{tab:gammaMAE}, together with Figures~\ref{fig:MAEratio2} and ~\ref{fig:MAEratio1}, we can see that QP\textsubscript{$in$} outperforms the KD\textsubscript{$rect$}, KD\textsubscript{$triw$} and PC estimators for almost every quantile, often by a wide margin. The performance differences are most pronounced on the right tail for the exponential example and both the left and right tails for the gamma example. For example, at the $p=0.99$ quantile in Table \ref{tab:gammaMAE}, the $MAE$ for QP\textsubscript{$in$} is about five times smaller than that for KD\textsubscript{$rect$} and six times smaller than that for KD\textsubscript{$triw$} and PC. This indicates that the undesirable tail oscillation illustrated in Fig.~\ref{fig:introplot} for the KD method is substantially mitigated by the QP method. The advantage of the QP method is more obvious when adding the additional shape constraints, especially for the exponential examples. The performance of the QP estimators with additional shape constraints, i.e. QP\textsubscript{$incms$} in Table~\ref{tab:expMAE} and QP\textsubscript{$incu$} in Table~\ref{tab:gammaMAE}, is as much as seven to eight times better than the KD and PC estimators for some quantiles. See Sec. \ref{sec:discuss_shape} for details on these and other constraints for the gamma and exponential examples. The only situations for which QP\textsubscript{$in$} did not outperform all other methods are for $p=0.5$ and $p=0.75$ for the gamma example in Table \ref{tab:gammaMAE}, where KD\textsubscript{$rect$} slightly outperformed QP\textsubscript{$in$}. Overall, the KD and PC methods are comparable with each other, with PC being slightly better than KD for the exponential example, which has a sharper pdf, and slightly worse than KD for the relatively smooth gamma example.

\begin{table}[htbp]
	\centering
	\caption{Comparison of quantile estimation $MAE$s ($\times10^3$) for various probabilities ($p$) with automated selection of regularization parameters for the exponential example.}
	\begin{tabular}{c|ccccc}
		\toprule
		$p$  & QP\textsubscript{$in$}& QP\textsubscript{$incms$}   & KD\textsubscript{$rect$} & KD\textsubscript{$triw$}& PC\\
		\midrule
		0.01  & 5.29  & 1.72  & 6.42  & 6.74  & 6.35 \\
		0.05  & 28.3  & 8.24  & 31.3  & 33.1  & 30.9 \\
		0.1   & 53.3  & 15.5  & 60.5  & 64.4  & 59.4 \\
		0.25  & 106   & 31.0  & 134   & 147   & 129 \\
		0.5   & 112   & 36.8  & 190   & 230   & 174 \\
		0.75  & 69.4  & 23.3  & 155   & 225   & 143 \\
		0.9   & 87.1  & 23.6  & 169   & 196   & 175 \\
		0.95  & 82.1  & 26.5  & 191   & 188   & 175 \\
		0.99  & 83.1  & 25.1  & 174   & 180   & 171 \\		
		\bottomrule
	\end{tabular}%
	\label{tab:expMAE}%
\end{table}%

\begin{table}[htbp]
	\centering
	\caption{Comparison of quantile estimation $MAE$s ($\times10^3$) for various probabilities ($p$) with automated selection of regularization parameters for the gamma example.}
	\begin{tabular}{c|ccccc}
		\toprule
		$p$  & QP\textsubscript{$in$}& QP\textsubscript{$incu$} & KD\textsubscript{$rect$} & KD\textsubscript{$triw$}& PC \\
		\midrule
		0.01  & 8.23  & 7.92  & 13.1  & 37.4  & 10.4 \\
		0.05  & 13.4  & 13.2  & 19.9  & 49.6  & 19.8 \\
		0.1   & 12.2  & 12.0  & 18.2  & 46.5  & 20.3 \\
		0.25  & 8.18  & 7.99  & 9.19  & 18.5  & 12.7 \\
		0.5   & 11.8  & 11.4  & 10.8  & 27.8  & 13.5 \\
		0.75  & 7.82  & 7.43  & 7.01  & 47.2  & 10.6 \\
		0.9   & 7.55  & 6.94  & 10.5  & 37.1  & 16.9 \\
		0.95  & 4.84  & 4.26  & 16.4  & 27.8  & 19.9 \\
		0.99  & 2.58  & 2.07  & 12.9  & 16.2  & 16.3 \\
		\bottomrule
	\end{tabular}%
	\label{tab:gammaMAE}%
\end{table}%

To eliminate any adverse effects of selecting the regularization parameter inappropriately and focus on the inherent performances of the approaches, we repeat the preceding MC simulations, but we use the oracle regularization parameters instead of automated parameter selection. The corresponding $MAE$s are listed in Tables~\ref{tab:expMAEoracle} and \ref{tab:gammaMAEoracle}. 
Since the PC method implemented by the \textbf{R} package \textbf{deamer} does not have a user-specified regularization parameter, the $MAE$s for the PC column listed in Tables~\ref{tab:expMAEoracle} and \ref{tab:gammaMAEoracle} are the same as those in Tables~\ref{tab:expMAE} and \ref{tab:gammaMAE}, although we still include it in Tables~\ref{tab:expMAEoracle} and \ref{tab:gammaMAEoracle} for comparison purposes. The $MAE$ ratio comparisons are displayed in Figures~\ref{fig:MAEratio4} and ~\ref{fig:MAEratio3}.

The oracle $\lambda$ for each example is determined by trying a broad range of $\lambda$ values and choosing the one that results in the smallest aggregate measure of $MAE$ across all nine quantiles. The aggregate measure is $\sum_{i=1}^9 MAE_i/[p_i(1-p_i)]$, where $MAE_i$ is the $MAE$ over the MC simulation for the quantile corresponding to $p_i$. The oracle $h$ values for KD\textsubscript{$rect$} and KD\textsubscript{$triw$} are determined similarly. Notice that the $MAE$s of KD\textsubscript{$rect$} are identical in Tables \ref{tab:expMAE} and \ref{tab:expMAEoracle} to three significant digits. This is because the KD\textsubscript{$rect$} bandwidth selection method in the exponential example is consistent across all MC replicates, and the selected bandwidth is quite close to the oracle $h$. Specifically, for the exponential example, the oracle regularization parameter values for QP\textsubscript{$in$}, QP\textsubscript{$incms$}, KD\textsubscript{$rect$} and KD\textsubscript{$triw$} are selected at 0.00995, 0.316, 0.720, 0.457, respectively. As for the gamma example, the oracle regularization parameter values for QP\textsubscript{$in$}, QP\textsubscript{$incu$}, KD\textsubscript{$rect$} and KD\textsubscript{$triw$} are selected at 0.011, 0.016, 0.867, 0.467, respectively.

\begin{table}[htbp]
	\centering
	\caption{Comparison of quantile estimation $MAE$s ($\times 10^3$) for various probabilities ($p$) with \textsl{oracle regularization parameters} for the exponential example.}
	\begin{tabular}{c|ccccc}
		\toprule
		$p$   & QP\textsubscript{$in$}& QP\textsubscript{$incms$} & KD\textsubscript{$rect$} & KD\textsubscript{$triw$}& PC\\
		\midrule
		0.01   & 5.78  & 1.71  & 6.42  & 6.56  & 6.35 \\
		0.05   & 27.6  & 8.19  & 31.3  & 32.1  & 30.9 \\
		0.1    & 51.6  & 15.4  & 60.5  & 62.1  & 59.4 \\
		0.25   & 101   & 31.2  & 134   & 139   & 129 \\
		0.5    & 107   & 36.1  & 190   & 209   & 174 \\
		0.75   & 67.6  & 22.2  & 155   & 200   & 143 \\
		0.9    & 85.7  & 24.2  & 169   & 179   & 175 \\
		0.95   & 78.7  & 27.2  & 191   & 174   & 175 \\
		0.99   & 82.0  & 24.7  & 174   & 168   & 171 \\
		
		\bottomrule
	\end{tabular}%
	\label{tab:expMAEoracle}%
\end{table}%

\begin{table}[htbp]
	\centering
	\caption{Comparison of quantile estimation $MAE$s ($\times 10^3$) for various probabilities ($p$) with \textsl{oracle regularization parameters} for the gamma example.}
	\begin{tabular}{c|ccccc}
		\toprule
		$p$  & QP\textsubscript{$in$} & QP\textsubscript{$incu$} & KD\textsubscript{$rect$} & KD\textsubscript{$triw$}& PC \\
		\midrule
		0.01   & 5.71  & 5.60  & 12.8  & 30.6  & 10.4 \\
		0.05   & 11.1  & 11.0  & 19.4  & 41.1  & 19.8 \\
		0.1    & 10.4  & 10.4  & 17.6  & 38.4  & 20.3 \\
		0.25   & 7.78  & 7.70  & 8.95  & 14.3  & 12.7 \\
		0.5    & 9.64  & 9.60  & 10.5  & 23.3  & 13.5 \\
		0.75   & 7.31  & 7.20  & 6.83  & 38.4  & 10.6 \\
		0.9    & 6.09  & 5.89  & 10.5  & 31.3  & 16.9 \\
		0.95   & 4.47  & 4.05  & 16.1  & 24.1  & 19.9 \\
		0.99   & 2.51  & 2.04  & 11.9  & 15.9  & 16.3 \\
		\bottomrule
	\end{tabular}%
	\label{tab:gammaMAEoracle}%
\end{table}%

Comparing Tables \ref{tab:expMAEoracle} and \ref{tab:gammaMAEoracle} with Table \ref{tab:expMAE} and \ref{tab:gammaMAE}, all methods perform only slightly worse in terms of $MAE$ performance when the regularization parameters are chosen automatically, relative to when the oracle values are used. Together with the results in Section \ref{sec:screeplot}, 
this implies that $\lmds$ is reasonably selected for most replicates, and the primary drawback of the automated SURE-like method is underestimation of $\lambda$ on a relatively small percentage of replicates (which can be easily corrected using the scree-plot method discussed in Section \ref{sec:screeplot}).

\subsection{The Effects of Incorporating Shape Constraints on the QP Estimator}\label{sec:discuss_shape}
The performance advantages of incorporating relevant shape constraints into the pdf estimation can be gauged from Tables~\ref{tab:expMAE}--\ref{tab:gammaMAEoracle}. In this section, we investigate this in more depth. Table \ref{tab:shape.effect} demonstrates the performance improvement that can be achieved by including various shape constraints in the QP method for the same gamma and exponential examples. Each number in Table \ref{tab:shape.effect} is the QP $MAE$ using the indicated constraint, divided by the QP $MAE$ using only the retro-$in$ constraints. The QP\textsubscript{$in$} $MAE$ values were the same as those shown in Tables \ref{tab:expMAE} and \ref{tab:gammaMAE}. An $MAE$ ratio less than 1.0 indicates that the $MAE$ is better when the constraints are included. For the exponential example, we use the support constraint $x\geq0$, and the monotonicity ($m$) and convexity constraints are over the entire support. For the gamma example, the $MAE$s for the left and right tails are substantially improved using the $in$ constraints versus using the retro-$in$ constraints, although little further improvement is achieved by using the additional convexity ($c$) and unimodality ($u$) constraints. We also considered the additional support ($s$) constraint of $x\geq0$, but the $incus$ constraints resulted in virtually identical performance as the $incu$ constraints for the gamma example and are omitted here. For the convexity constraint for the gamma example, we take the left and right tails to be convex beyond the inflection points for the true pdf of $X$, which in practice would have to be approximated roughly by inspection of the histograms. The inflection points are $x=2$ and $x=6$ for the $Gamma(5,1)$ pdf. For the unimodality constraint, the mode location is treated as unknown and is automatically estimated by the QP method.

\begin{table}[htbp]
	\centering
	\caption{Effect of adding shape constraints on the performance of QP. Each number in the table is the ratio of the quantile $MAE$ using the indicated constraints, divided by the corresponding quantile $MAE$ using retro-$in$.\label{tab:shape.effect}}
	\begin{tabular}{c|ccc|cc}
		\toprule
		&\multicolumn{3}{c|}{Exponential} & \multicolumn{2}{c}{Gamma} \\
		$p$  & \textit{in} & \textit{incm} & \textit{incms}&\textit{in} & \textit{incu} \\
		\midrule
		0.01  & 0.925 & 0.466 & 0.269 & 0.494 & 0.475  \\
		0.05  & 0.909 & 0.465 & 0.264 & 0.612 & 0.601 \\
		0.1   & 0.889 & 0.466 & 0.258 & 0.662 & 0.655 \\
		0.25  & 0.818 & 0.472 & 0.239 & 1.027 & 1.004 \\
		0.5   & 0.634 & 0.489 & 0.208 & 0.928 & 0.890  \\
		0.75  & 0.518 & 0.493 & 0.173 & 1.134 & 1.078 \\
		0.9   & 0.619 & 0.443 & 0.168 & 1.136 & 1.044 \\
		0.95  & 0.544 & 0.454 & 0.176 & 0.924 & 0.814 \\
		0.99  & 0.564 & 0.447 & 0.170 & 0.497 & 0.398 \\
		\bottomrule
	\end{tabular}%
\end{table}

We can see from Table~\ref{tab:shape.effect} that for the exponential example, the $MAE$s for every quantile are improved by incorporating the $in$ constraints, substantially so for the middle quantiles and right tail. Moreover, incorporating the additional $cm$ and then $s$ constraints successively lead to substantial further improvement in the $MAE$s.  For example, for the upper quantiles of the exponential distribution, the $MAE$ is almost an order of magnitude smaller using the $incms$ constraints, relative to using only the retro-$in$ constraints.

\subsection{Performance Comparisons in Terms of PDF Estimation} \label{sec:discuss_pdf}

The results in the previous sections focused on quantile estimation, which is closely related to cdf estimation. If the goal of the data analysis is to produce numerical characteristics of the distribution of $X$, then cdf and quantile estimation will usually be more relevant than pdf estimation. However, since a plot of the estimated pdf can be an useful graphical complement to the numerical cdf and quantile values and provide insight into the distribution of $X$, a direct assessment of the pdf estimation is also of interest. In this section we compare the pdf estimators for the QP, KD and PC methods. 

In Table \ref{tab:exp_pdf_decomp_oracle} and Table~\ref{tab:gamma_pdf_decomp_oracle}, we compare the QP, KD and PC pdf estimators in terms of their bias, standard deviation (SD), and root mean square error (RMSE) for the estimated pdf at nine different quantiles. Note that $\text{RMSE}^2 = \text{Bias}^2 + \text{SD}^2$.  The numbers displayed in each cell of Table \ref{tab:exp_pdf_decomp_oracle} and Table~\ref{tab:gamma_pdf_decomp_oracle} are in format Bias/SD/RMSE. Overall, the QP estimators perform better than the other methods, and the difference is even larger when incorporating more reasonable shape constraints into the QP estimators. In particular, for the exponential example, the additional shape constraints beyond $in$ improve the performance of the QP method substantially. In contrast, for the gamma example, the additional shape constraints only help a little, and only at the upper quantiles. Similar conclusions about the effect of incorporating additional shape constraints were made for the cdf comparisons in Tables~\ref{tab:expMAEoracle} and \ref{tab:gammaMAEoracle}.

\begin{table}[htbp]
	\centering
	\caption{MC comparison of the Bias/SD/RMSE ($\times 10^2$) for the pdf estimators at quantiles corresponding to various probabilities ($p$) for the exponential example using oracle regularization parameters.}
	\begin{tabular}{c|ccccc}
		\toprule
		$p$  & QP\textsubscript{$in$} & QP\textsubscript{$incms$} & KD\textsubscript{$rect$} & KD\textsubscript{$triw$}& PC \\
		\midrule
		0.01&  -25.5 /25.6 /36.2& -7.36 /7.48 /10.5& -28.2 /28.3 /40.0& -28.9 /28.9 /40.9& -28.3 /28.5 /40.2\\
		0.05&  -22.2 /22.3 /31.5& -6.79 /6.90 /9.68& -25.7 /25.8 /36.4& -26.5 /26.5 /37.5& -25.5 /25.7 /36.2\\
		0.1 &  -18.1 /18.2 /25.7& -5.68 /5.80 /8.12& -22.6 /22.6 /32.0& -23.4 /23.5 /33.2& -22.0 /22.1 /31.2\\
		0.25&  -7.05 /7.52 /10.3& -2.49 /2.63 /3.62& -13.5 /13.5 /19.1& -14.6 /14.7 /20.7& -12.0 /12.2 /17.1\\
		0.5 &  -0.70 /2.88 /2.96&  0.66 /0.85 /1.08& -2.49 /2.88 /3.81&  3.61 /4.32 /5.63& -0.37 /2.84 /2.87\\
		0.75&  -0.10 /2.28 /2.29&  0.54 /0.86 /1.01&  2.26 /2.38 /3.28&  1.59 /2.09 /2.63& -0.73 /3.82 /3.89\\
		0.9 &  -0.07 /1.82 /1.82& -0.35 /0.54 /0.64& -2.40 /2.47 /3.44&  0.49 /1.17 /1.26& -0.76 /2.51 /2.63\\
		0.95&   0.25 /1.32 /1.35& -0.03 /0.29 /0.29& -0.09 /0.43 /0.44&  0.21 /0.87 /0.89&  0.69 /1.68 /1.82\\
		0.99&   0.01 /0.52 /0.52&  0.01 /0.13 /0.13& -0.44 /0.44 /0.62&  0.06 /0.37 /0.38&  0.42 /1.14 /1.22\\
		\bottomrule
	\end{tabular}%
	\label{tab:exp_pdf_decomp_oracle}%
\end{table}%

\begin{table}[!tbp]
	\centering
	\caption{MC comparison of the Bias/SD/RMSE ($\times 10^2$) for the pdf estimators at quantiles corresponding to various probabilities ($p$) for the gamma example using oracle regularization parameters.}
	\begin{tabular}{c|ccccc}
		\toprule
		$p$  & QP\textsubscript{$in$} & QP\textsubscript{$incu$} & KD\textsubscript{$rect$} & KD\textsubscript{$triw$}& PC \\
		\midrule
0.01&  1.02 /1.26 /1.62 & 1.04 /1.25 /1.63 & 1.55 /1.66 /2.27 & 2.48 /2.63 /3.62 & 1.16 /2.28 /2.56\\
0.05&  0.17 /0.86 /0.88 & 0.18 /0.88 /0.90 & 0.15 /0.64 /0.66 & 0.27 /1.03 /1.07 &-0.32 /2.39 /2.41\\
0.1 & -0.68 /1.16 /1.34 &-0.70 /1.16 /1.35 &-0.94 /1.12 /1.46 &-1.42 /1.78 /2.28 &-1.35 /2.92 /3.21\\
0.25& -1.19 /1.50 /1.91 &-1.21 /1.51 /1.93 &-1.69 /1.77 /2.45 &-3.21 /3.40 /4.67 &-1.99 /3.72 /4.22\\
0.5 &  0.26 /1.01 /1.05 & 0.27 /1.01 /1.04 &-0.17 /0.70 /0.72 &-2.12 /2.41 /3.21 &-0.96 /3.61 /3.74\\
0.75&  0.54 /1.05 /1.18 & 0.53 /1.03 /1.16 & 0.21 /0.78 /0.81 &-0.02 /1.10 /1.10 &-0.60 /2.90 /2.96\\
0.9 & -0.32 /0.83 /0.89 &-0.31 /0.80 /0.85 &-0.70 /0.92 /1.16 & 0.62 /1.10 /1.26 &-0.42 /2.00 /2.05\\
0.95& -0.18 /0.66 /0.69 &-0.15 /0.56 /0.58 &-0.29 /0.58 /0.65 & 0.51 /0.91 /1.04 & 0.00 /1.75 /1.75\\
0.99&  0.04 /0.38 /0.38 & 0.04 /0.25 /0.26 & 0.23 /0.45 /0.50 & 0.19 /0.50 /0.53 & 0.26 /1.31 /1.33\\
\bottomrule
\end{tabular}%
	\label{tab:gamma_pdf_decomp_oracle}%
\end{table}%

When we look at the quantile (and cdf by proxy, since it is closely related to quantiles) and pdf comparisons together, we see an interesting phenomenon regarding the relative performance improvement of QP\textsubscript{$in$} over KD\textsubscript{$rect$}. Namely, the performance improvement for quantile estimation appears to be much greater than for pdf estimation. For example, from Table~\ref{tab:exp_pdf_decomp_oracle}, the pdf RMSE values for QP\textsubscript{$in$} are overall better than for KD\textsubscript{$rect$} for the exponential example (QP\textsubscript{$in$} is substantially better at some $p$ values, comparable at others, and substantially worse at one value $p=0.95$). In contrast, from Table~\ref{tab:expMAEoracle}, the quantile estimation results for QP\textsubscript{$in$} are substantially better than for KD\textsubscript{$rect$} at nearly every $p$ value and worse at none. Similarly, for the gamma example, the pdf RMSE performances (Table~\ref{tab:gammaMAEoracle}) of QP\textsubscript{$in$} and KD\textsubscript{$rect$} are comparable overall, whereas the quantile performance (Table~\ref{tab:gamma_pdf_decomp_oracle}) for QP\textsubscript{$in$} is substantially better than for KD\textsubscript{$rect$}.

\begin{figure}[!tbp]
  \centering
  {\includegraphics[width=0.75\textwidth]{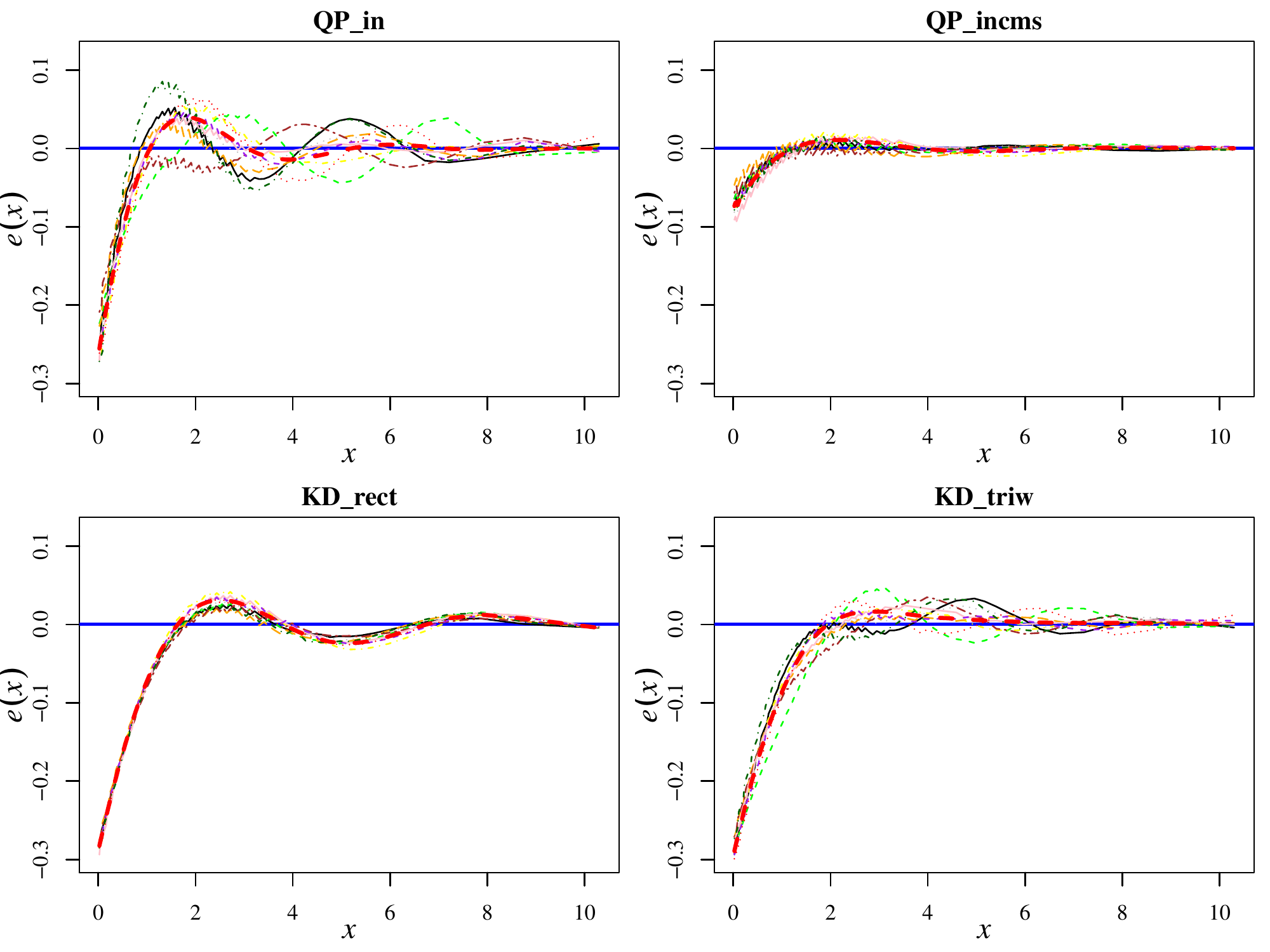}}
   \caption{Plots of $e(x)=\hat{f}(x)-f(x)$ vs. $x$ for 10 typical replicates for the exponential example for four different methods.}\label{fig:experror}
\end{figure}

\begin{figure}[!tbp]
  \centering
  {\includegraphics[width=0.75\textwidth]{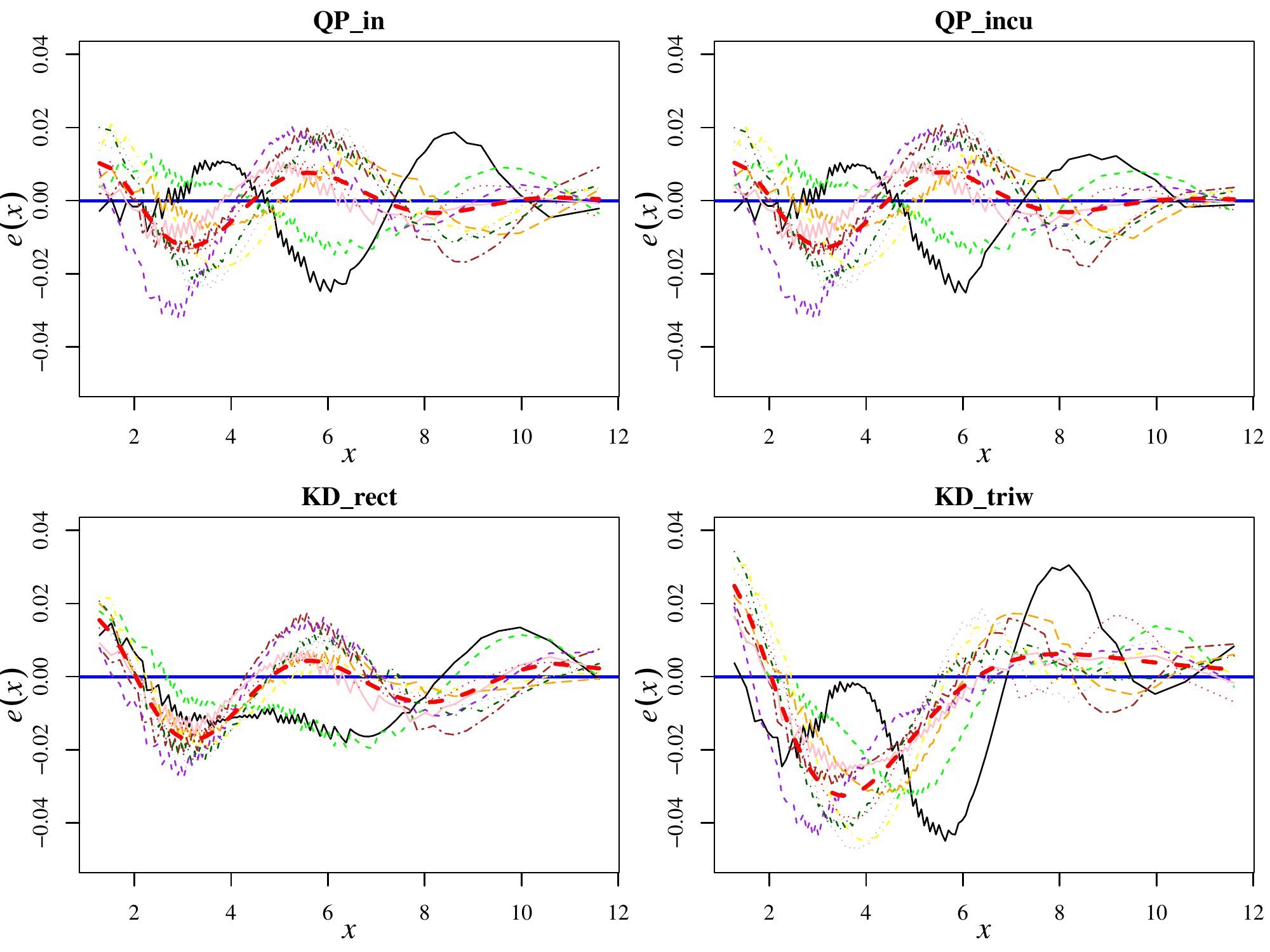}}
  \caption{Plots of $e(x)=\hat{f}(x)-f(x)$ vs. $x$ for 10 typical replicates for the gamma example for four different methods.}\label{fig:gammaerror}
\end{figure}

One possible explanation is that the pdf estimation error $e(x) \equiv \hat{f}(x)-f(x)$ for the QP pdf estimator at different $x$ tends to be less positively (or more negatively) correlated than for the KD estimator, so that when one integrates the pdf to compute the cdf, the errors tend not to accumulate as much for the QP estimator. To investigate this, Figures~\ref{fig:experror} and~\ref{fig:gammaerror} plot the error $e(x)$ as a function of $x$ for the QP and KD estimators for 10 representative MC replicates of the above exponential and gamma examples, respectively. The thick blue solid horizontal line indicates $e(x)=0$, and the thick red dashed curve shows the mean function for $e(x)$ (the mean is taken pointise in $x$ across all MC replicates). From Figures~\ref{fig:experror} and~\ref{fig:gammaerror}, it appears that for each individual replicate, the error function $e(x)$ for the QP estimator does tend to oscillate more than for the KD estimator, in the sense that the $e(x)$ curves have more zero-crossings for the QP estimator. 

\begin{figure}[!tbp]
  \centering
  {\includegraphics[width=0.8\textwidth]{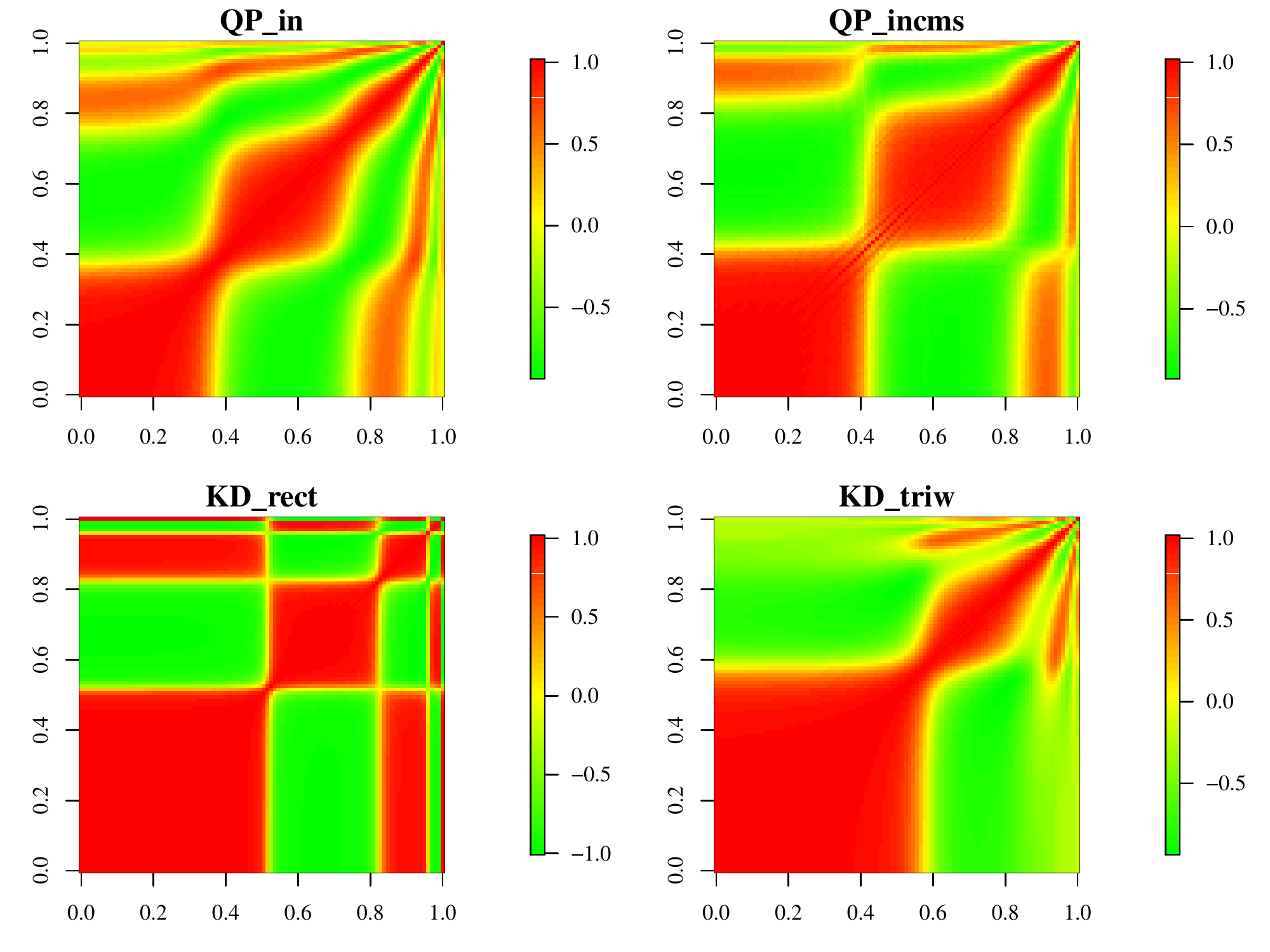}}
  \caption{Heatmaps of the uncentered correlations across all MC replicates for the exponential example for four different methods.}\label{fig:correxperror}
\end{figure}

\begin{figure}[!tbp]
  \centering
  {\includegraphics[width=0.8\textwidth]{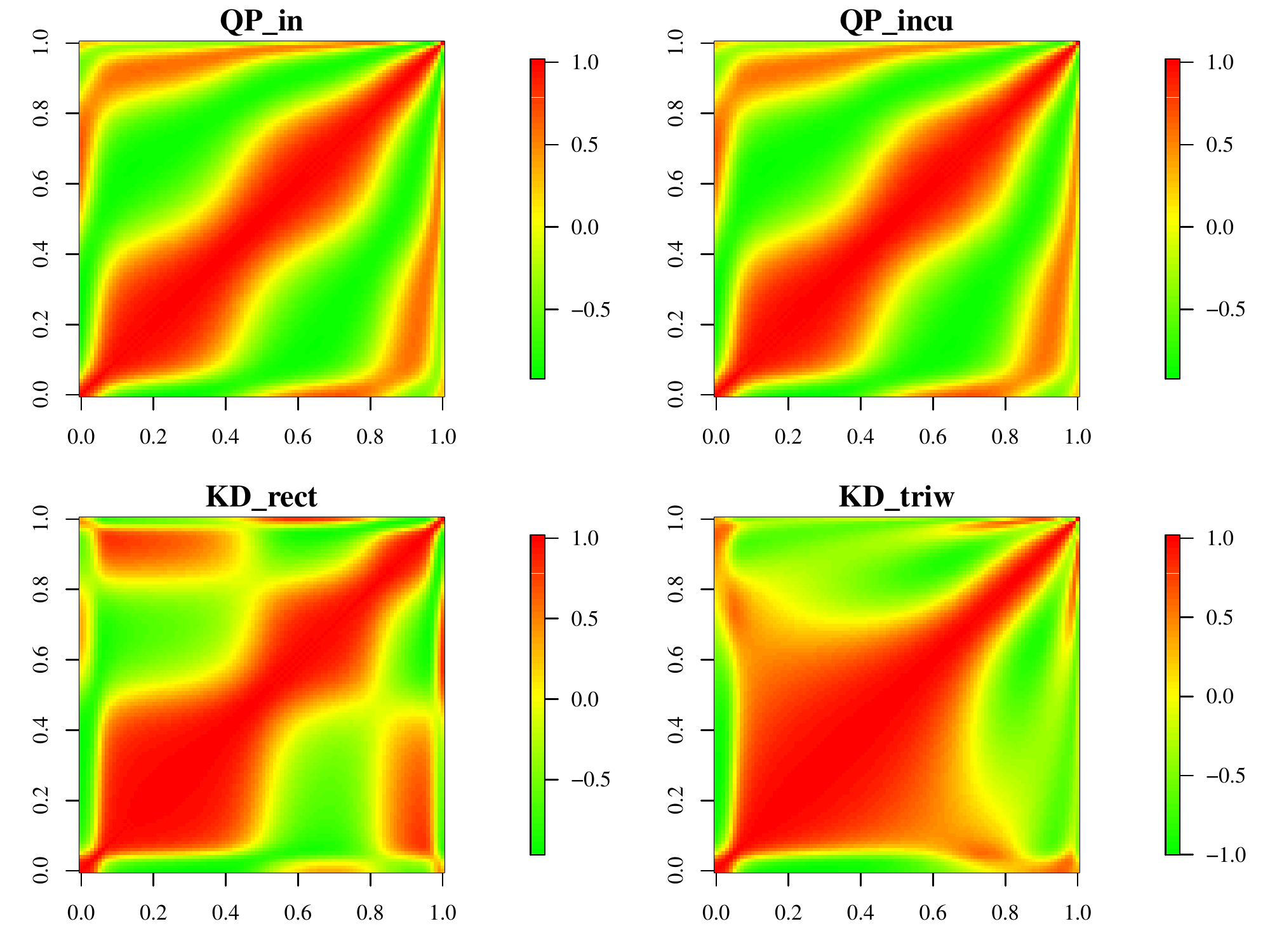}}
  \caption{Heatmaps of the uncentered correlations across all MC replicates for the gamma example for four different methods.}\label{fig:corrgammaerror}
\end{figure}

To further investigate this, Figures~\ref{fig:correxperror} and~\ref{fig:corrgammaerror} show heatmaps of the uncentered correlation matrix for the pdf error function $e(x)$ across all MC replicates for the exponential and the gamma examples, respectively. To obtain the uncentered correlation matrix, we first construct the $K\times N$ error matrix $\mathrm{\bm{V}}^T = [e_1(\x),e_2(\x),\cdots,e_N(\x)]$, where $e_i^T(\x)= \hat{f}(\x)-f(\x)$ is the $1\times K$ error vector for the $i$th MC replicate, $\x$ is a $K$-length vector of evenly-spaced values covering the domain of the true pdf, and $N$ denotes the number of MC replicates. The uncentered correlation matrix is defined as $\mathrm{\bm{A}}^{-1/2}\mathrm{\bm{V}}^T\mathrm{\bm{V}}\mathrm{\bm{A}}^{-1/2}/N$, where $\mathrm{\bm{A}} = \text{diag}(\mathrm{\bm{V}}^T\mathrm{\bm{V}}/N)$. Unlike the usual correlation matrix, the uncentered correlation matrix detects association in both systematic and random variation. From Figures~\ref{fig:correxperror} and~\ref{fig:corrgammaerror}, it appears that the band of strong positive correlation along the diagonal is stronger for KD\textsubscript{$rect$} than for QP\textsubscript{$in$}. The fact that the positive correlation in $e(x)$ at nearby $x$ values is stronger for KD\textsubscript{$rect$} than for QP\textsubscript{$in$} implies that when the estimated pdf is integrated to obtain the estimated cdf, the estimation errors will tend to accumulate more for KD\textsubscript{$rect$}.

\section{Conclusions}\label{sec:conclusion}
In this article, we have developed and investigated a new method of density deconvolution, based on quadratic programming (QP) with constraints, for the additive measurement error model. The method enjoys substantially better deconvolution performance than existing methods across the examples that we considered. We have also developed an \textbf{R} package \textbf{QPdecon} to implement the approach. 

Our QP method appears to have a more favorable tradeoff between oversmoothing versus tail oscillation than other density deconvolution methods like the KD and the wavelet-like PC methods. 
Via the examples in Section \ref{sec:discussion}, we have demonstrated that the QP method with only the universally applicable $in$ (integrate-to-one and nonnegativity) constraints can perform substantially better than the KD and PC methods, especially at estimating the tail quantiles of the distribution (see Tables \ref{tab:expMAE} and \ref{tab:gammaMAE} and Fig.~\ref{fig:MAEratio}). Furthermore, a major advantage of the QP deconvolution method is that a number of frequently relevant density constraints (unimodality, tail monotonicity, tail convexity, support constraints) can be easily incorporated directly into the density estimation. For examples like the exponential one, including additional constraints dramatically improved the performance (see Table \ref{tab:shape.effect}). 

Our QP method contains two schemes to select the regularization parameter. The first scheme is the automated SURE-like method, and the second scheme is a graphical scree-plot method. For a relatively small percentage of replicates, the automatically selected $\lmds$ is unreasonably small, which results in an erratic $\hfx$ like the one shown in Fig.~\ref{fig:fig3b}. This occurred on approximately 5\% of the replicates for the gamma example and 1\% of the replicates for the exponential example. However, this can be remedied using the simple but effective scree-plot method, as illustrated in Fig.~\ref{fig:screeplot}. In the examples that we have considered, all of the outlier replicates having the largest estimation error in $\hfx$ were caused by underestimation of $\lmds$, and this was readily apparent via simple inspection of the scree plot. Moreover, in these situations the scree plot also suggested a better choice for $\lambda$ that resulted in a much better $\hfx$.

Throughout, we have computed the QP cdf estimator (from which the quantiles are calculated) simply by integrating the QP pdf estimator. An alternative would be to reformulate the QP deconvolution method to work directly with the cdfs, instead of the current formulation that works with the pdfs. We have explored this alternative cdf formulation and found that the resulting direct QP cdf estimator was no better than integrating the QP pdf estimator. We also found that QP pdf estimator was a much better pdf estimator than discrete-differencing the direct QP cdf estimator. Consequently, we have only discussed the pdf formulation.

We have focused on the density deconvolution estimator itself, as opposed to attempting to quantify the uncertainty in the estimator. Regarding the latter, bootstrapping methods could be used if desired. 
However, to conserve space and focus on new ideas, we will not discuss bootstrapping in this paper. Assessing the bias in the estimator would be difficult, which is true for any density deconvolution estimator that involves some form of regularization, including the KD and PC methods. 

\section*{Acknowledgements}
This work was supported in part by NSF Grant CMMI-1436574, which the authors gratefully acknowledge.

\bibliographystyle{abbrvnat}
\bibliography{main}

\section{Appendix}
\subsection[Derivation of Eq. (8)]{Derivation of Eq.~(\ref{Errexp})}
Recall that $\hfy$ and $\hfy^0$ are independent random vectors with a common mean $\C\fx$. We can view $\C\hfxl$, with $\hfxl$ the estimator from the QP approach, to be an estimator of this common mean. We write the SURE criterion as:
\begin{eqnarray}\label{A1}
\E[Err]&=&\E\left[\lVert\hfy^0-\C\hfxl\rVert^2\right] \nonumber\\
&=&\E\left[\lVert(\hfy^0-\C\fx)-(\hfy-\C\fx)+(\hfy-\C\hfxl)\rVert^2\right] \nonumber\\
&=&\E\left[\lVert\hfy^0-\C\fx\rVert^2\right]+\E\left[\lVert \hfy-\C\fx\rVert^2\right]+\E\left[\lVert\hfy-\C\hfxl\rVert^2\right]\nonumber\\
&&-2\E\left[(\hfy^0-\C\fx)^T(\hfy-\C\fx)\right]+2\E\left[(\hfy^0-\C\fx)^T(\hfy-\C\hfxl)\right]\nonumber\\
&&-2\E\left[(\hfy-\C\fx)^T(\hfy-\C\hfxl)\right].			
\end{eqnarray}
Term by term in (\ref{A1}), we have $\E\left[\lVert\hfy^0-\C\fx\rVert^2\right]=\E\left[\lVert\hfy-\C\fx\rVert^2 \right]$ (because $\hfy^0$ is defined as a random draw from the same distribution as $\hfy$); $\E\left[\lVert\hfy-\C\hfxl\rVert^2 \right]=\E[err]$ (by definition of err); and $\E\left[(\hfy^0-\C\fx)^T(\hfy-\C\fx)\right]=\E\left[(\hfy^0-\C\fx)^T(\hfy-\C\hfxl)\right]=0$ (by the independence of $\hfy^0$ and $\hfy$). Consequently, (\ref{A1}) reduces to:
\begin{eqnarray*}
	\E[Err]&=&\E[err]+2\left\{\E\left[(\hfy-\C\fx)^T(\hfy-\C\fx)\right]-\E\left[(\hfy-\C\fx)^T(\hfy-\C\hfxl)\right]\right\}\\
	&=&\E[err]+2\E\left[(\hfy-\C\fx)^T(\C\hfxl-\C\fx)\right]\\
	&=&\E[err]+2\mathrm{tr}\left[\COV(\C\hfxl,\hfy)\right].
\end{eqnarray*}
The last equality follows even if $\E[\hfxl]\neq\fx$, because $\hfy-\C\fx$ is zero-mean.

\end{document}